\documentclass[a4paper,fleqn]{cas-sc}
\usepackage[numbers]{natbib}
\usepackage[]{lineno}
\def\tsc#1{\csdef{#1}{\textsc{\lowercase{#1}}\xspace}}
\tsc{WGM}
\tsc{QE}
\begin{document}
\let\WriteBookmarks\relax
\def\floatpagepagefraction{1}
\def\textpagefraction{.001}

%Short title
\shorttitle{INTT Silicon Ladder for sPHENIX}    

\shortauthors{<Y. Akiba, {\it et al.}>}  

\title[mode=title]{The Ladder and Readout Cables of Intermediate Silicon Strip Detector for sPHENIX}  

\tnotemark[1] 

\tnotetext[1]{This work is supported by } 


\author[1,2]{Y.~Akiba}[]
\author[7,1]{H.~Aso}[]
\author[9,2]{J.~T.~Bertaux}[]
\author[3]{D.~Cacace}[]
\author[4]{K.~Y.~Chen}[]
\author[4,2]{K.~Y.~Cheng}[]
\author[1]{A.~Enokizono}[]
\author[1,2]{H.~Enyo}[]
\author[7,1]{K.~Fujiki}[]
\author[7,1]{Y.~Fujino}[]
\author[5,1]{M.~Fujiiwara}[]
\author[5,2]{T.~Hachiya}[]
\author[7,1]{T.~Harada}[]
\author[6]{S.~Hasegawa}[]
\author[5,2]{M.~Hata}[]
\author[11]{B.~Hong}[]
\author[11,2]{J.~Hwang}[]
\author[7,1]{T.~Ichino}[]
\author[5,1]{M.~Ikemoto}[]
\author[7,1]{H.~Imagawa}[]
\author[7,1,2]{H.~Imai}[]
\author[5,1]{Y.~Ishigaki}[]
\author[5]{M.~Isshiki}[]
\author[5,1]{K.~Iwatsuki}[]
\author[5]{R.~Kan}[]
\author[5,1]{M.~Kano}[]
\author[7,1]{T.~Kato}[]
\author[7,1]{R.~Kawashima}[]
\author[7,1]{T.~Kikuchi}[]
\author[8]{T.~Kondo}[]
\author[4]{C.~M.~Kuo}[]
\author[5]{H.~Kureha}[]
\author[1]{T.~Kumaoka}[]
\author[9]{H.~S.~Li}[]
\author[10]{R.~S.~Lu}[]
\author[3]{E.~Mannel}[]
\author[7,1]{H.~Masuda}[]
\author[2]{G.~Mitsuka}[]
\author[5,1]{N.~Morimoto}[]
\author[5,2]{M.~Morita}[]
\author[1,2]{I. Nakagawa}[]\cormark[1]%\ead{itaru@riken.jp}
\author[7,1]{Y.~Nakamura}[]
\author[7,1]{G.~Nakano}[]
\author[5,2]{Y.~Namimoto}[]
\author[7,1]{D.~Nemoto}[]
\author[5]{S.~Nishimori}[]
\author[3]{R.~Nouicer}[]
\author[1]{G.~Nukazuka}[]
\author[5,1]{I.~Omae}[]
\author[3]{R.~Pisani}[]
\author[1]{Y.~Sekiguchi}[]
\author[5,2]{M.~Shibata}[]
\author[4,2]{C.~W.~Shih}[]
\author[7,1]{K.~Shiina}[]
\author[5]{M.~Shimomura}[]
\author[7,1]{R.~Shishikura}[]
\author[9,2]{M.~Stojanovic}[]
\author[5]{K.~Sugino}[]
\author[5]{Y.~Sugiyama}[]
\author[5,2]{A.~Suzuki}[]
\author[5,2]{R.~Takahama}[]
\author[10]{L.~S.~Tsai}[]
\author[4,2]{W.~C.~Tang}[]
\author[5]{Y.~Terasaka}[]
\author[2]{T.~Todoroki}[]
\author[5,1]{H.~Tsujibata}[]
\author[7,1]{T.~Tsuruta}[]
\author[2]{Y.~Yamaguchi}[]
\author[7,1]{H.~Yanagawa}[]
\author[5,2]{M.~Watanabe}[]
\author[9]{R.~Xiao}[]
\author[9]{W.~Xie}[]




%\ead[url]{http://riken.jp}

%\credit{<Credit authorship details>}

\affiliation[1]{organization={Nishina Center for Accelerator-Based Science, RIKEN},
            addressline={2-1 Hirosawa}, 
            city={Wako},
            postcode={351-0198}, 
            state={Saitama},
            country={Japan}}

\affiliation[2]{organization={RIKEN BNL Research Center},
            addressline={20 Pennsylvania Avenue}, 
            city={Upton},
            postcode={11973}, 
            state={NY},
            country={U.S.A.}}            
            
\affiliation[3]{organization={Brookhaven National Laboratory},
            addressline={20 Pennsylvania Avenue}, 
            city={Upton},
            postcode={11973}, 
            state={NY},
            country={U.S.A.}}

\affiliation[4]{organization={Department of Physics and Center for High Energy and High Field Physics, National Central University},
            addressline={No.300, Zhongda Rd., Zhongli Dist.}, 
            city={Taoyuan City},
            postcode={32001}, 
            country={Taiwan}}

\affiliation[5]{organization={Department of Mathematical and Physical Sciences, Nara Women's University},
            addressline={Kitauoya-Higashimachi}, 
            city={Nara},
            postcode={630-8506}, 
            state={Nara},
            country={Japan}}         
            
\affiliation[6]{organization={Advanced Science Research Center, Japan Atomic Energy Agency},
            addressline={2-4 Shirakata Shirane}, 
            city={Tokai-mura, Naka-gun},
            postcode={319-1195}, 
            state={Ibaraki},
            country={Japan}}
            
\affiliation[7]{organization={Rikkyo University, Department of Physics},
             addressline={3-34-1 Nishi-Ikebukuro, Toshima}, 
            city={Tokyo},
            postcode={171-8501}, 
            state={Tokyo},
            country={Japan}} 
            
\affiliation[8]{organization={Tokyo Metropolitan Industrial Technology Research Institute},
             addressline={2-4-10, Aomi, Koto}, 
            city={Tokyo},
            postcode={ 135-0064}, 
            state={Tokyo},
            country={Japan}} 

\affiliation[9]{organization={Department of Physics and Astronomy, Purdue University},
             addressline={525 Northwestern Ave.}, 
            city={West Lafayette},
            postcode={47907}, 
            state={IN},
            country={U.S.A.}} 
            
\affiliation[10]{organization={Department of Physics, National Taiwan University},
             addressline={No.1 Sec.4 Roosevelt Road}, 
            city={Taipei},
            postcode={10617}, 
            country={Taiwan}} 
            
\affiliation[11]{organization={Korea University, Department of Physics},
             addressline={Anam-dong 5, Seongbuk-gu}, 
            city={Seoul},
            postcode={02841}, 
            country={Korea}} 
            
% Footnote of the second author

% Email id of the second author
%\ead{}



\cortext[1]{I.~Nakagawa}

%\fntext[1]{xxxx}


\begin{abstract}
A new silicon-strip-type detector was developed for precise charged-particle tracking in the central rapidity region of heavy ion collisions. A new detector and collaboration at the Relativistic Heavy Ion Collider at Brookhaven National Laboratory is sPHENIX, which is a major upgrade of the PHENIX detector. The intermediate tracker (INTT) is part of the advanced tracking system of the sPHENIX detector complex together with a CMOS monolithic-active-pixel-sensor based silicon-pixel vertex detector, a time-projection chamber, and a micromegas-based detector. The INTT detector is barrel shaped and comprises 56 silicon ladders. Two different types of strip sensors of 78~$\mu m$ pitch and 320~$\mu m$ thick are mounted on each half of a silicon ladder. Each strip sensor is segmented into 8$\times$2 and 5$\times$2 blocks with lengths of 16 and 20 mm. Strips are read out with a silicon strip-readout (FPHX) chip. In order to transmit massive data from the FPHX to the down stream readout electronics card (ROC), a series of long and high speed readout cables were developed. This document focuses on the silicon ladder, the readout cables, and the ROC of the INTT. The radiation hardness is studied for some parts of the INTT devices in the last part of this document, since the INTT employed some materials from the technology frontier of the industry whose radiation hardness is not necessarily well known.
\end{abstract}


%\item INTT
%\item sPHENIX
%\item RHIC

% Keywords

\maketitle

% Main text

%%%%%%%%%%%%%%%%%%%%%%%%%%%%%%%%%%%%%%%%%%%%%%%%%%
%%%%%%%%%%%%%%%% Detecor Overview %%%%%%%%%%%%%%%%
%%%%%%%%%%%%%%%%%%%%%%%%%%%%%%%%%%%%%%%%%%%%%%%%%%
\section{Introduction}
\label{Introduction}

The sPHENIX detector~\cite{Aidala:2012nz} at the Relativistic Heavy Ion Collider (RHIC) at Brookhaven National Laboratory, USA, is a major upgrade of the PHENIX detector~\cite{Adcox:2003nim}, which was decommissioned in 2017. The sPHENIX experiment collects high-statistics proton-proton, proton-nucleus, and nucleus-nucleus data, enabling state-of-the-art studies of jet modification, upsilon suppression, and open heavy-flavor production to probe the microscopic nature of the strongly-coupled quark-gluon plasma. Such measurements are complementary to those of experiments at the Large Hadron Collider (LHC) at CERN, and will also allow a broad range of cold quantum chromodynamic studies~\cite{Belmont:2023fau}. 

\begin{figure}[htb]
\begin{center}
\includegraphics[scale=0.40]{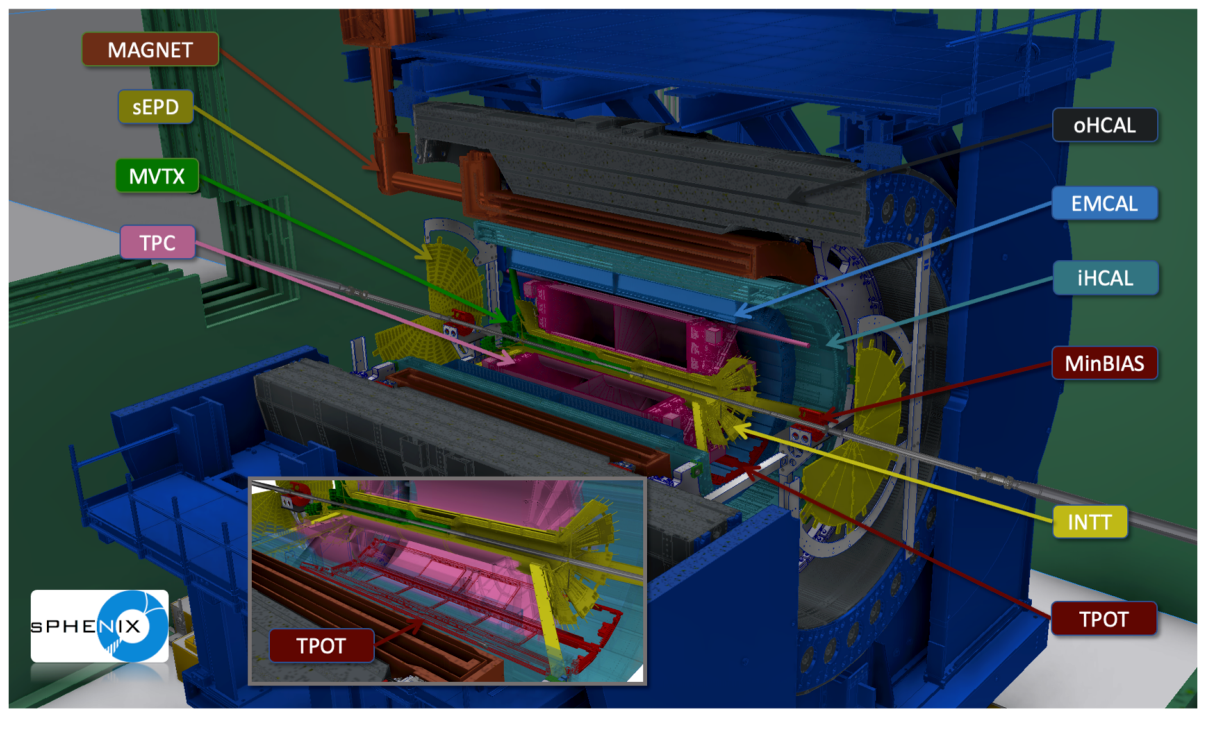}
\caption{
The mechanical drawing of the sPHENIX detector.
\label{fig:sPHENIX}
}
\end{center}
\end{figure}

The sPHENIX detector provides precision vertexing, tracking, and electromagnetic and hadronic calorimetry in the central pseudorapidity region $|\eta|$<1.1 with full azimuthal coverage at the full RHIC collision rate of 15 kHz in Au+Au at $\sqrt{s}=200$ GeV. A comprehensive assessment of these requirements led to the development of the reference design shown in Fig.~\ref{fig:sPHENIX}. In its overall layout, sPHENIX follows the typical geometry of modern collider detectors. The tracking system comprises a CMOS monolithic-active-pixel-sensor (MAPS) microvertex detector (MVTX), a silicon-strip intermediate tracker (INTT), and a time-projection chamber (TPC). For calibration, a micromegas based detector (TPOT)~\cite{Aune:2024nim} partially covers the outside of the TPC acceptance. 
The calorimeter stack includes a tungsten/scintillating fiber electromagnetic calorimeter (EMCAL) and aluminum/scintillator and steel/scintillator tile hadronic calorimeter (HCAL), divided into inner and outer parts. The inner HCAL sits inside the 1.4~T superconducting solenoid, which was refurbished from the decommissioned BaBar detector~\cite{Aubert2002nim}.

This document describes the details of the INTT silicon ladder, its readout cables, and the readout electronics card. 

%%%%%%%%%%%%%%%%%%%%%%%%%%%%%%%%%%%%%%%%%%%%%%%%%%
%%%%%%%%%%%%%%%% Detecor Overview %%%%%%%%%%%%%%%%
%%%%%%%%%%%%%%%%%%%%%%%%%%%%%%%%%%%%%%%%%%%%%%%%%%
\section{Detector Overview}
\label{Overview}
The barrel-type INTT detector comprises inner and outer layers of INTT silicon ladders. Adjacent ladders are staggered to prevent dead space in azimuthal acceptance.  The inner and outer barrels have 24 and 32 ladders, respectively, as summarized in Table~\ref{tbl:INTT_Barrel}. Details of the INTT barrel are beyond the scope of this document, and a separate paper is under preparation. 

\begin{table}[h]
\begin{center}
\caption{Number of the INTT silicon ladders of the inner and outer layers of the INTT barrel detector.}
\label{tbl:INTT_Barrel}
\begin{tabular}{l|c|c}
\hline\hline
                  & Inner & Outer\\
\hline
Number of ladders &  24   &  32 \\
\hline\hline
\end{tabular}
\end{center}
\end{table}

The designs of the INTT ladder and its readout are constrained by the specification of of the FPHX~\cite{PHENIX_FPHX_cite, Kapustinsky:2010nim, FPHX_Manual} readout chip and the readout card (ROC) of the Forward Vertex (FVTX) detector~\cite{Aidala:2013vna} in PHENIX. Since the INTT reuses the FPHX design and ROCs, the rest of INTT detector was designed to be compatible with these readout chip and electronics. Thus, new copies of FPHX chips were produced for the INTT detector, while the ROCs were recycled. The major difference between the FVTX and the INTT detectors is in the mechanical concept. The INTT is a barrel type with 2 layers, while the FVTX was a disk type with 4 layers.

\begin{figure}[h]
	\centering
		\includegraphics[scale=0.4]{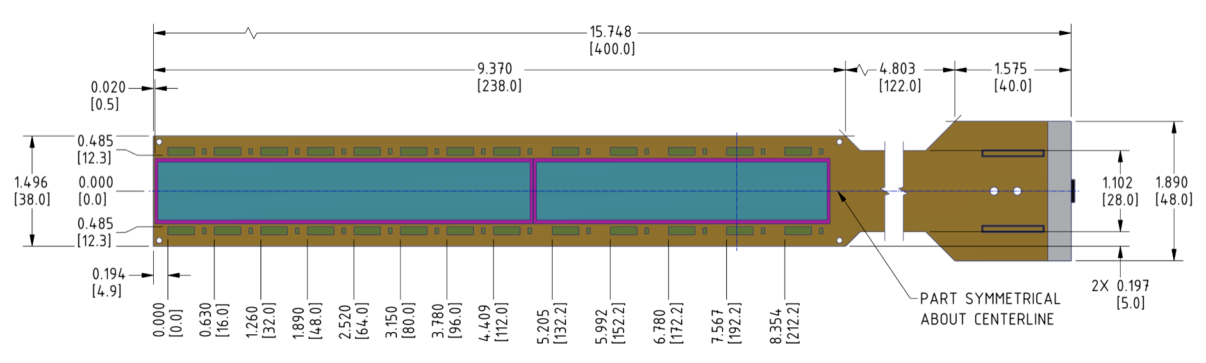}
	  \caption{Schematics of the INTT module and the dimensions of each components, i.e. the silicon sensors, FPHX chips, and the HDI cable. Dimensions are given in inches, while numbers in brackets are in millimeters.} \label{INTT_SilionOnHDI_Drawing}
\end{figure}

Two silicon sensors and 26 FPHX readout chips~\cite{Aidala:2013vna, Kapustinsky:2010nim, FPHX_Manual} are mounted on a high-density interconnect (HDI) flexible print cable to form an INTT silicon module. The schematics of the silicon module are shown in Fig.~\ref{INTT_SilionOnHDI_Drawing}. The layout of the silicon sensors and the FPHX chips with respect to the HDI cable is defined in the figure.

\begin{figure*}[h]
	\centering
		\includegraphics[scale=0.8]{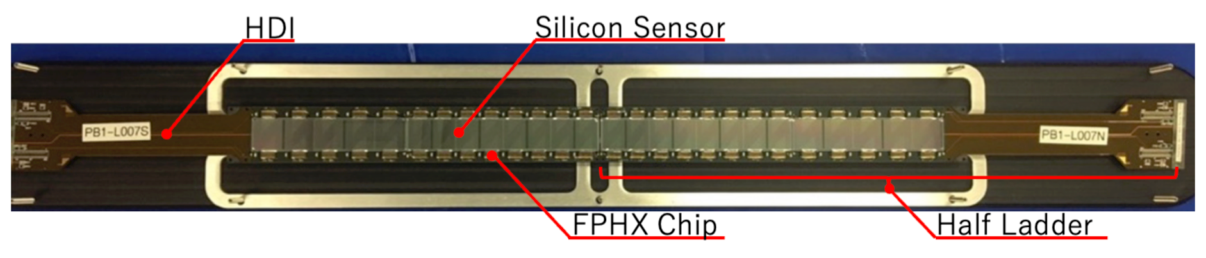}
	  \caption{Photo of the INTT ladder with sensors facing up. Note the center line dividing the two halves of the sensor and the rows of FPHX chips along the sensor edges.} \label{INTT_Ladder_photo}
\end{figure*}

\begin{figure}[h]
	\centering
		\includegraphics[scale=0.4]{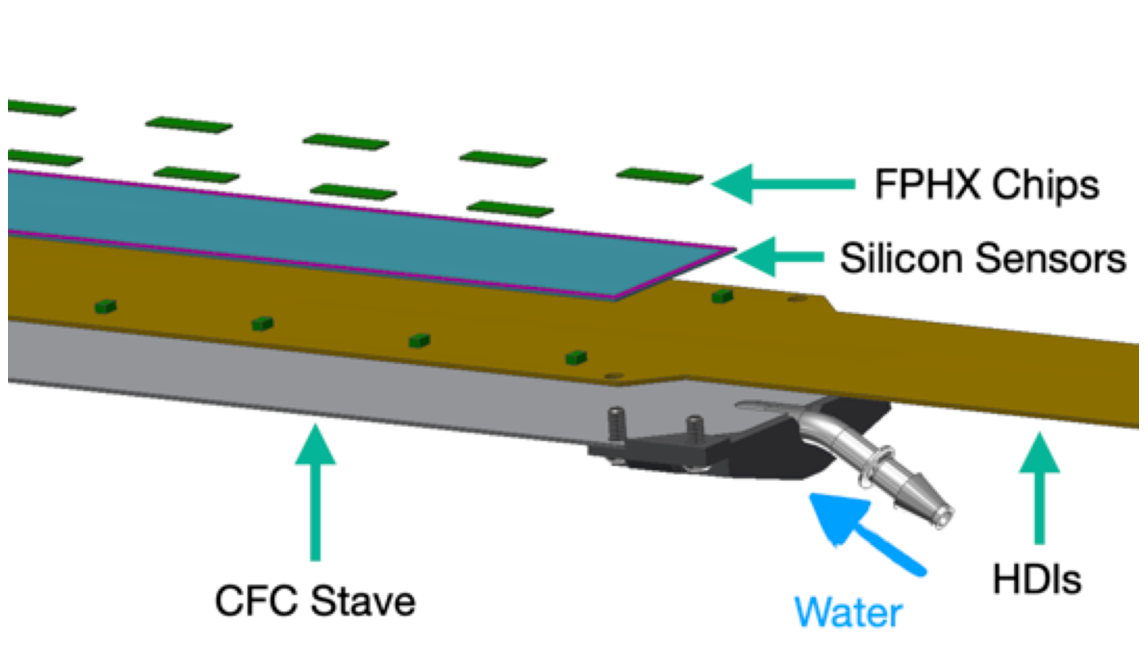}
	  \caption{The stackup of the components of the INTT ladder.} \label{INTT_LadderStructure}
\end{figure}

These components of the INTT silicon module and the ladder are summarized in Table~\ref{tbl:LadderComponents}. Two INTT modules are aligned longitudinally and glued to a carbon fiber composite (CFC) support stave to form an INTT silicon ladder, as shown in Fig.~\ref{INTT_Ladder_photo}. 
To show the structures of the INTT ladder, the stackup of the components of the ladder is illustrated in Fig.~\ref{INTT_LadderStructure}.

Both glue and carbon fiber have high thermal conductivity to diffuse heat generated by the FPHX chips. A water cooling system removes heat from the barrel through a carbon tube integrated within the body of the stave. Figure~\ref{INTT_LadderStructure} also shows the tube attached to the edge of the CFC stave, which is an inlet for cooling water.

\begin{table}[h]
\begin{center}
\caption{The components and total number of channels of the INTT module and ladder. The details of silicon strip sensor type-A and B are discussed in the section\ref{Silicon}. 
\label{tbl:LadderComponents}}
\begin{tabular}{l|c|c}
\hline\hline
               & Component                &  Quantity\\
\hline
\multirow{4}{4em}{Silicon module} & Silicon strip sensor     & 2 (Type-A \& B) \\
               & FPHX Chips               & 26 \\
               & HDI                      & 1 \\
               & Total number of channels & 3,328\\
\hline
\multirow{3}{4em}{Silicon ladder}& Silicon module           & 2 \\
               & Stave                    & 1 \\
               & Total number of channels & 6,656\\
\hline\hline
\end{tabular}
\end{center}
\end{table}

The data of a given ladder are read out from both longitudinal ends for each silicon module. Thus, each HDI cable transmits data from half of a ladder. 
The data are further transmitted downstream from the HDI by a bus extender (BEX) cable~\cite{BEX} and a conversion cable (CC). The BEX is a 1.11 meter long flexible print cable (FPC) employing liquid-crystal polymer as a dielectric material to suppress losses in transmission lines. In both ends, the conversion cable comprises three components; 1) $\mu$-coaxial harness, 2) power and ground cables, and 3) connector print boards. The downstream end of the conversion cable is connected to the read-out card (ROC) which collects data from up to seven half-ladders, and then transmits reformatted data to further downstream electronics via an optical fiber connection. The downstream electronics beyond the ROC and data acquisition system are beyond the scope of this document.

%%%%%%%%%%%%%%%%%%%%%%%%%%%%%%%%%%%%%%%%%%%%%%%%%%
%%%%%%%%%%%%%%%% INTT Ladder %%%%%%%%%%%%%%%%%%%%%
%%%%%%%%%%%%%%%%%%%%%%%%%%%%%%%%%%%%%%%%%%%%%%%%%%
\section{INTT Ladder}
\label{INTT Ladder}
This section describes the electrical components and support systems used to read out and power the INTT. The silicon-strip sensors and the FPHX readout chips are introduced in subsections~\ref{Silicon} and \ref{FPHX chip}, respectively. The HDI that provides power, bias voltage, and slow-control signals to the sensor is discussed in subsection~\ref{HDI}. The stave support structure is discussed in subection~\ref{Stave}. The material budget of the ladder in the radial direction is discussed in subsection~\ref{MaterialBudget}. Lastly, the dead space of the silicon ladder is evaluated in subsection~\ref{Dead Space}.

%%%%%%%%%%%%%%%%%%%%%%%%%%%%%%%%%%%%%%%%%%%%%%%%%%%%%%%%%%%%%%%
\subsection{Silicon Sensors}
%%%%%%%%%%%%%%%%%%%%%%%%%%%%%%%%%%%%%%%%%%%%%%%%%%%%%%%%%%%%%%%
\label{Silicon}

The silicon-strip sensor~\cite{SiliconHPK} is single-sided and AC-coupled.  The design is based on and modified from the PHENIX FVTX silicon mini-strip sensors~\cite{Aidala:2013vna}. The sensors (model S14629-01), as well as the FVTX sensors, were fabricated by Hamamatsu Photonics K.K. There are two types of sensors (type-A and type-B), which are distinguished by the length of strips and number of blocks. The type-A sensor has an active area of 128~$\times$~19.968 mm that is segmented into 8~rows and 
2~columns of blocks, each of which consisted of 128 strips with 78~$\mu$m pitch and 16~mm long. Each strip oriented in the longitudinal direction.  Similarly, the type-B sensor has an active area of 100~$\times$~19.968 mm that is segmented into 5~rows and 2~columns of blocks, each of which comprises 128 strips in 78~$\mu$m pitch and 20~mm long that are also oriented in the longitudinal direction. Table~\ref{tbl:SiliconDemensions} summarizes these dimensions.

\begin{table*}[h]
\begin{center}
\caption{The dimensions of type-A and type-B silicon strip sensors~\cite{SiliconHPK}.}
\label{tbl:SiliconDemensions}
\begin{tabular}{l|c|c|c}
\hline\hline
Item & Type-A & Type-B & Type-A+B Total \\
\hline
Physical dimensions & 130.0 mm $\times$ 22.5 mm & 102.0 mm $\times$ 22.5 mm & 232.0 mm $\times$ 22.5 mm \\
Active area dimensions & 128 mm $\times$ 19.968 mm & 100 mm $\times$ 19.968 mm & 228 mm  $\times$ 19.968 mm \\
Active area fraction & 87.4\% & 87.0\% &  \\
\hline
Active cell dimensions & 16 mm $\times$ 19.968 mm & 20 mm $\times$ 19.968 mm & N/A \\
\# of cells  &  8 &  5 & 13 \\
\hline
Active block dimensions & 16 mm $\times$ 9.984 mm & 20 mm $\times$ 9.984 mm & N/A \\
\# of blocks & 16 & 10 & 26 \\
\hline
Number of blocks per cell & \multicolumn{3}{c}{2}  \\
Number of strips per block & \multicolumn{3}{c}{128}  \\
Strip pitch &  \multicolumn{3}{c}{78 $\mu$m}  \\
\hline
Number of strip channels &  2,048 & 1,280 & 3,328 \\
\hline\hline
\end{tabular}
\end{center}
\end{table*}

\begin{figure}[htb]
\begin{center}
\includegraphics[scale=0.55]{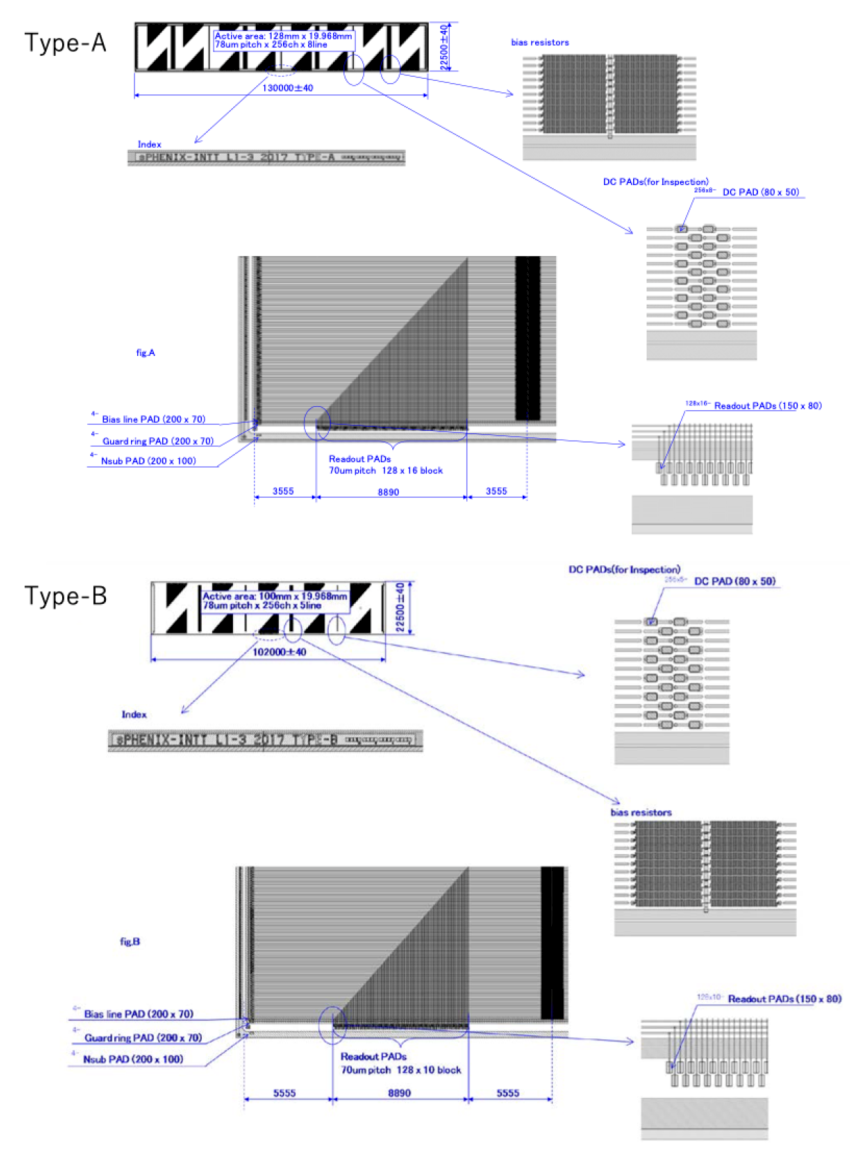}
\caption{
The layout design of the type-A (top) and type-B (bottom) silicon sensors~\cite{SiliconHPK}. 
\label{fig:INTT_SiliconLayout}
}
\end{center}
\end{figure}

 The small gap between adjacent blocks where the DC pads are laid out on the surface is completely active. The sensors were fabricated with p-implants on a 320~$\mu$m thick n-type substrate. The strips are AC-coupled and biased through individual 15~M$\Omega$ poly-silicon resistors to a typical operating voltage of 100~V. The aluminum metallization width of the strips is 20~$\mu$m, which is wider than the implant width of 10~$\mu$m. This provides field-plate protection against micro discharges, which are known to grow with radiation-induced increases in the leakage current. These specifications of the silicon sensor are summarized in Table~\ref{tbl:SiliconSpecification}.
 The strips are also protected by two p-implant guard rings and an n+ region between the guard rings and the sensor edge. These designs follow from the silicon-strip sensors of the FVTX~\cite{Aidala:2013vna}. Figure~\ref{fig:INTT_SiliconLayout} shows details of the sensor layout, including guard rings, bond-pad locations, and mechanical fiducial marks.

\begin{figure}[htb]
\begin{center}
\includegraphics[scale=0.48]{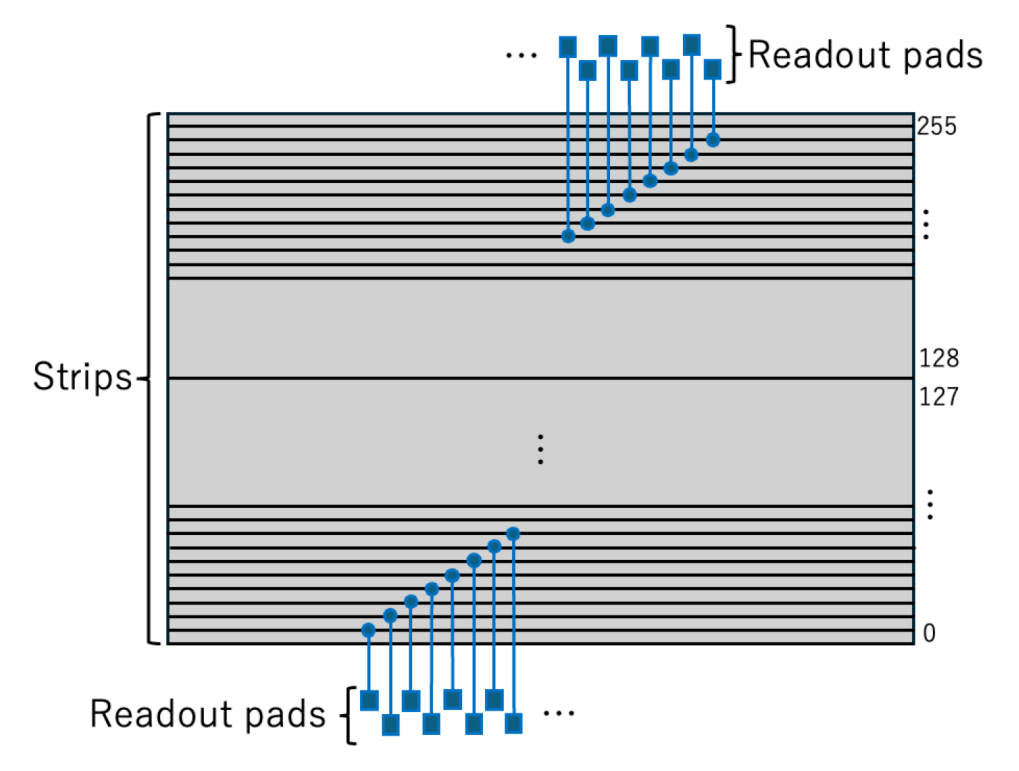}
\caption{
The schematics of the double metal structured strips and their readout lines~\cite{SiliconHPK}. 
\label{fig:INTT_Silicon_Schematic}
}
\end{center}
\end{figure}

 In Fig.~\ref{fig:INTT_Silicon_Schematic}, the strip runs horizontally (longitudinal direction). The readout lines of each strip are wired perpendicular to the strips orientation using double-metal technology. The other end of the readout lines are connected to readout pads, which transmit data to the FPHX chips using a wire bonding. The strips channel 0 to 127 are wired to the readout pads laid out on the bottom of the figure, while the strips from 128 to 255 are wired to the pads on top of the figure. 

\begin{table}[h]
\begin{center}
\caption{Specifications of the silicon strip sensors~\cite{SiliconHPK}.
\label{tbl:SiliconSpecification}}
\begin{tabular}{l|c}
\hline\hline
Item &  Specification \\
\hline
SSD type & AC-SSSD  \\
Nominal operating voltage  & 100 V\\
Bias Providing Type & Poly-Si bias\\
Poly-Si resistance  &  15 M$\Omega$\\
Silicon thickness & 320 $\mu$m \\
Strip implant width & 10 $\mu$m \\
Strip readout aluminum width  & 20 $\mu$m \\
\hline\hline
\end{tabular}
\end{center}
\end{table}

%%%%%%%%%%%%%%%%%%%%%%%%%%%%%%%%%%%%%%%%%%%%%%%%%%%%%%%%%%%%%%%
\subsection{FPHX chip}
%%%%%%%%%%%%%%%%%%%%%%%%%%%%%%%%%%%%%%%%%%%%%%%%%%%%%%%%%%%%%%%
\label{FPHX chip}

A custom 128-channel front-end ASIC, the FPHX~\cite{PHENIX_FPHX_cite, Kapustinsky:2010nim, FPHX_Manual}, was developed at Fermilab for use in the PHENIX FVTX Detector~\cite{Aidala:2013vna}. The size of the chip is 9~$\times$~2 mm. The chip is operated at 2.5~V and consumes power as low as 64~mW per chip. The FPHX is a mixed-mode chip with two major and distinct sections: the analog front-end, and the digital back-end.

The analog section consists of an integration/shaping stage, followed by a 3-bit ADC stage. The FPHX chip integrates and shapes signals from 128 channels of strips, digitizes and sparsifies the hit channels for each beam crossing, and serially reads out the digitized data.

The back-end is a novel, triggerless, data-push architecture that permits operation without deadtime and high-speed readout with very low latency. It has been designed to process up to four hits within four RHIC beam crossings. Although it takes longer, it can process more than 4 hits from a given crossing.

A fully processed hit pattern is zero suppressed, and contains a 7-bit timestamp in the unit of RHIC repetition frequency 9.4~MHz ($\sim 1/106$~ns), a 7-bit channel ID, and a 3-bit ADC value. The data word is output over two LVDS serial lines in alternating order at a rate of up to 200~MHz. 
A summary of FPHX specifications~\cite{PHENIX_FPHX_cite, Kapustinsky:2010nim, FPHX_Manual} is given in Table~\ref{tbl:FPHXSpecification}.

\begin{table}[h]
\begin{center}
\caption{Specifications of the FPHX readout chip~\cite{PHENIX_FPHX_cite, Kapustinsky:2010nim, FPHX_Manual}.
\label{tbl:FPHXSpecification}}
\begin{tabular}{l|c}
\hline\hline
Item &  Specification \\
\hline
Dimensions & 9 mm $\times$ 2 mm \\
Operation voltage &  2.5 V \\
Power consumption & 64 mW \\
Number of Channels & 128 \\
ADC channels & 3 bits \\
Data Transmission & 200 MHz \\
\hline\hline
\end{tabular}
\end{center}
\end{table}

In addition, to be as self-sufficient as possible, the FPHX chip provides its own internal bias voltages and currents with minimal external support circuitry.  For the user to be able to control the internal parameters and biases, a digital slow-control interface is provided on each chip to enable programming.  Adjustable parameters include gain, threshold, rise time, fall time, input transistor bias current, channel masking, and several additional fine-tuning parameters~\cite{FPHX_Manual}.

%%%%%%%%%%%%%%%%%%%%%%%%%%%%%%%%%%%%%%%%%%%%%%%%%%%%%%%%%%%%%%%
\subsection{High-density interconnects}
%%%%%%%%%%%%%%%%%%%%%%%%%%%%%%%%%%%%%%%%%%%%%%%%%%%%%%%%%%%%%%%
\label{HDI}

The HDI is a flexible print circuit board used to read out a half ladder, which comprises two silicon sensors with 26 FPHX chips. The basic layer design of the HDI structure follows from the FVTX~\cite{Aidala:2013vna}. The geometric constraint for the silicon ladder is somewhat less stringent for the INTT than for the FVTX.  Thus, the circuit design parameters such as line and space are relaxed from the FVTX, which results in an improved yield rate in the fabricating process.

The HDI was designed by HAYASHI-REPIC Co., Ltd. and fabricated by YAMASHITA MATERIALS CORPORATION in Japan. 
The width of the HDI is 38~mm in the sensor area and 43~mm at the connector end. The length is 398~mm, which is the longest limit of industry fabrication for the multilayer FPC in an automated manner using dedicated fabrication machines. 

\begin{figure}[htb]
\begin{center}
\includegraphics[scale=0.50]{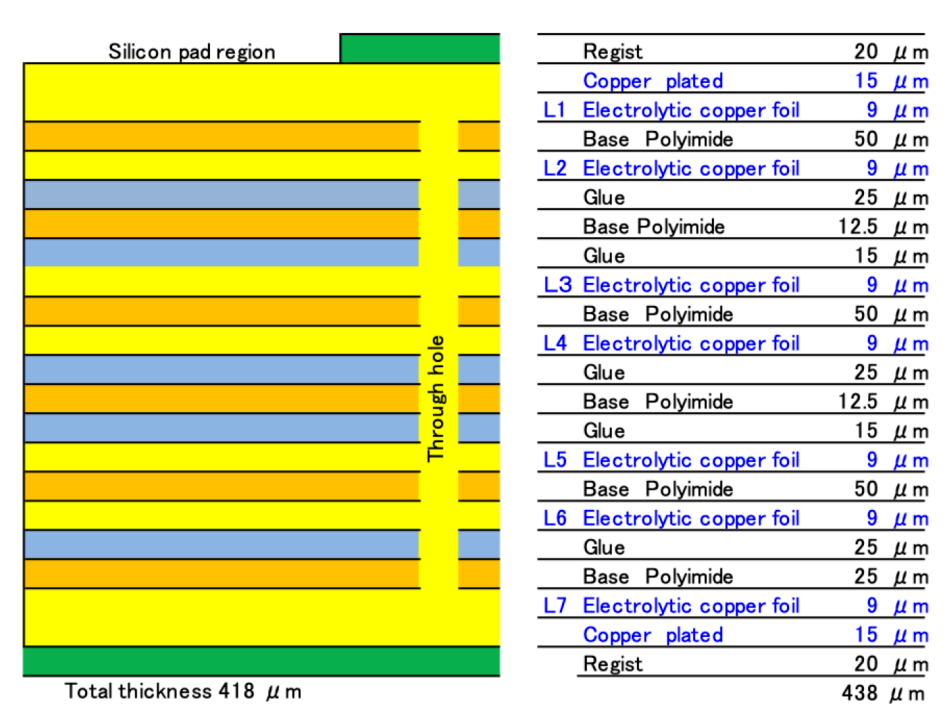}
\caption{
Seven layer structure of HDI. 
\label{fig:INTT_HDI_LayerCrossSection}
}
\end{center}
\end{figure}

The HDI is seven-layered, with each layer comprising 9~$\mu$m thick electrolytic copper foil, 12.5 to 50~$\mu$m thick polymide, and 15 to 25~$\mu$m thick resin glue. The total thickness is 418~$\mu$m in the sensor area.  As discussed in subsection~\ref{MaterialBudget}, the total thickness of copper layers is 68~$\mu$m, and is the greatest contribution to the material budget of the INTT ladder. The top and bottom copper foil layers were plated with 15~$\mu$m of copper\footnote{The copper plate was 15~$\mu$m thick for the 1st and 2nd batches of the HDI production. The thickness increased to 25~$\mu$m for the 3rd batch in HDI production. The total radiation length of the silicon ladder was increased by 3\%.}. The cross section of the seven-layer structure of the HDI is shown in Fig.~\ref{fig:INTT_HDI_LayerCrossSection}.

\begin{figure}[htb]
\begin{center}
\includegraphics[scale=0.40]{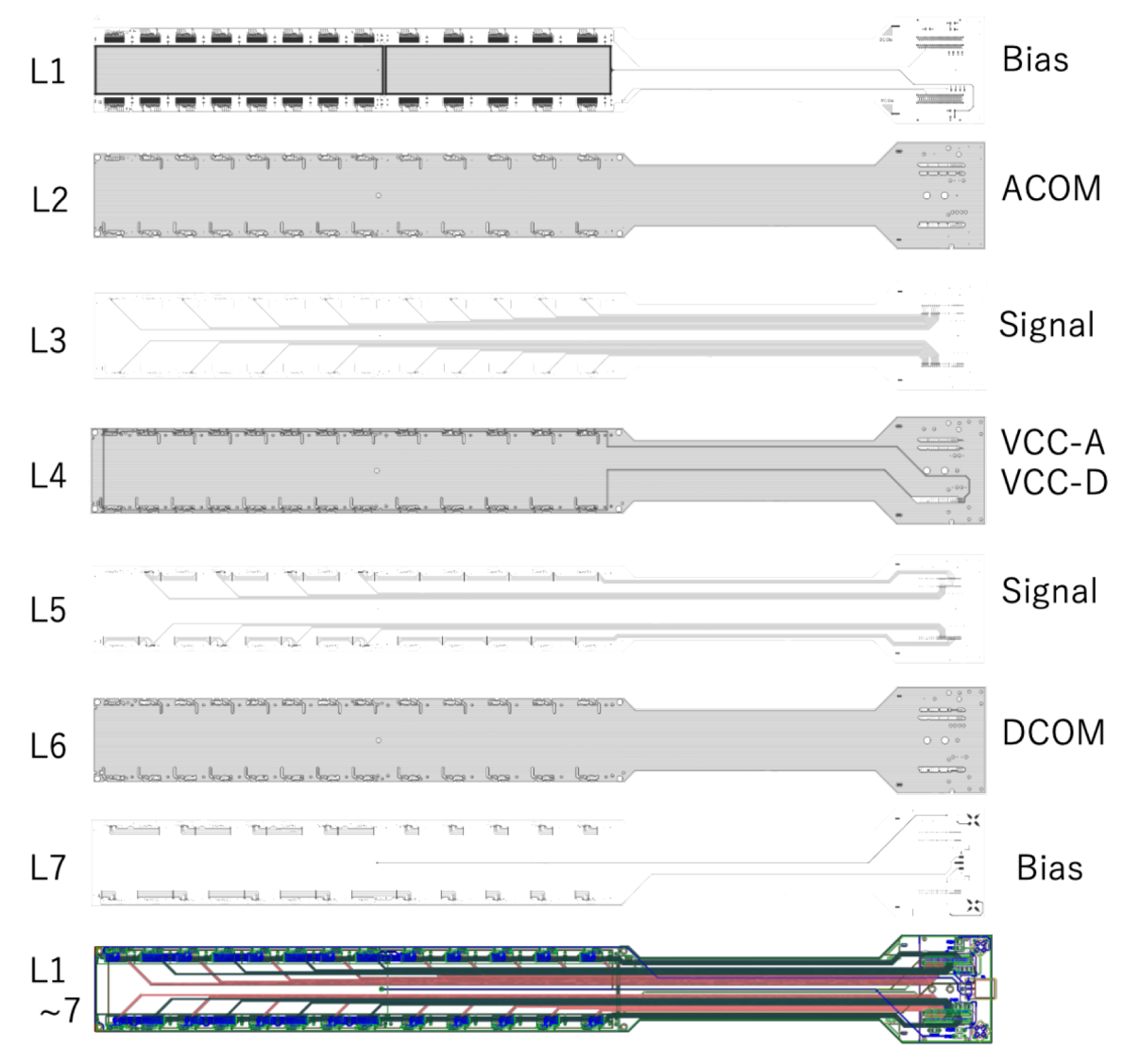}
\caption{
The schematic of each layer of the HDI. The ACOM and DCOM layers provide analog and digital grounds to FPHX chips, respectively. The 2.5~V power lines for the analog and digital parts of the FPHX chip are electrically isolated from each other within L4. The bottom schematic figure shows (in different colors) the overlay of line patterns for all seven layers.  
\label{fig:INTT_HDI_Layout}
}
\end{center}
\end{figure}

The schematic of each layer is shown in Fig.~\ref{fig:INTT_HDI_Layout}. From top to bottom, each layer is for: transmitting bias for the type-B silicon sensor (L1), analog ground (L2), signals (L3), DC power for the analog and digital parts of FPHX (L4), signal (L5), digital ground (L6), and for the type-A sensor and signals (L7). There are two thermostats implemented on the L7 to monitor the cooling status of heat generated during operation from the FPHX chips.  

There are 122 signal lines per HDI cable: 52 pairs of output data lines, 8 pairs of slow-control \& clock LVDS lines, and two dedicated lines to inject calibration pulses to the FPHX.  
Signal transmission lines are mainly placed in layers 3 and 5 to be shielded from external electromagnetic (EM) fields by keeping these layers between solid-copper layers which are assigned for either analog or digital grounds or the DC power. The line and space are both 60~$\mu$m. The characteristic impedance is designed to provide a 100 ${\rm \Omega}$ differential for these LVDS pair lines to ensure matching with the rest of the readout cable chain.

\begin{figure}[htb]
\begin{center}
\includegraphics[scale=0.4]{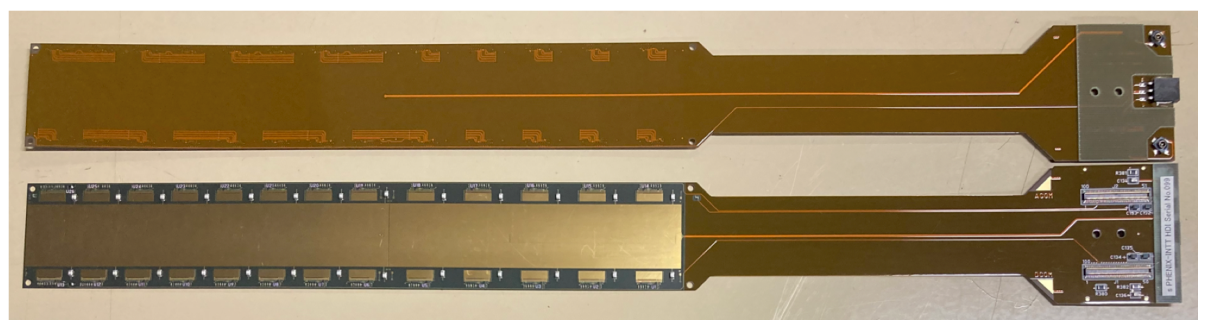}
\caption{
Photo of the HDI cables with L1 face up (bottom) and L7 face up (top). 
\label{fig:INTT_HDI_photo}
}
\end{center}
\end{figure}

Some signal lines in the sensor region L7 are not of the same design as in the FVTX. These signal lines were changed to keep the HDI width as narrow as possible and would not have otherwise fit within the signal layers. This narrow width enables the barrel detector to be as round as possible by minimizing physical interference between adjacent ladders. Because the back plane of L7 is not shielded by the solid foil ground nor the power layers, the signal lines are exposed to external EM fields. Thus, the length of the lines were kept as short as possible (centimeter lengthscale). Figure~\ref{fig:INTT_HDI_photo} shows a photo of the HDI cables with L1 and L7 face up. 
%There are some signal lines wired in the 7th layer which is not shielded from possible external electromagnetic field though, their lengths are short enough not to be severely suffered from noise.  
Table~\ref{tbl:HDISpecification} summarizes the major specifications of the HDI cable. 

\begin{table}[h]
\begin{center}
\caption{Specifications of the HDI cable. The characteristic impedance is for the LVDS pair.}
\label{tbl:HDISpecification}
\begin{tabular}{l|c}
\hline\hline
Item &  Specification \\
\hline
Dimensions  & 398 mm $\times$ 38 mm \\
Width of connector ends & 43 mm \\
Total thickness (sensor pad region) & 418 $\mu$m \\
Conductive material & Copper \\
Dielectric material & Polymide \\
Number of layers & 7 \\
Number of signal lines & 122 \\
Line and space &  60 $\mu$m \& 60 $\mu$m \\
Characteristic impedance & 100 ${\rm \Omega}$\\
\hline\hline
\end{tabular}
\end{center}
\end{table}

Pairs of connectors for data, LV power for the FPHX, and grounds are implemented in the end of the HDI. The model of data connector is the Hirose Co. DF18C-100DP-0.4V(51) plug, which consists of 100 channels with a 400~$\mu$m pitch between conducting pins. Also, there are two bias connectors and a 3-pin thermostat connectors implemented in the end of the HDI.

%%%%%%%%%%%%%%%%%%%%%%%%%%%%%%%%%%%%%%%%%%%%%%%%%%%%%%%%%%%%%%%
\subsection{Stave}
%%%%%%%%%%%%%%%%%%%%%%%%%%%%%%%%%%%%%%%%%%%%%%%%%%%%%%%%%%%%%%%
\label{Stave}

The stave is a mechanical support made mainly of carbon fiber composite (CFC) skins. The stave itself, plus an extension for mechanical attachment, spans 497~mm long and 38~mm wide. This matches the HDI width around the sensor area and accommodates two silicon modules per stave, forming the INTT silicon ladder.
The heat load expected from each half ladder is 64~mW~$\times$~26 FPHX chips, or approximately 1.7 W. The total heat load over the entire INTT is about 186~W for 112 half ladders. 

The stave is required not only for the rigidity of the support structure, but also for high thermal conductivity to dissipate local heat generated by the FPHX chips. 
The main component of the stave is made of a carbon fiber reinforced plastic prepreg (CFRP).  
The GRANOC prepreg sheet model NT91500-520S of Nippon Graphite Fiber Co. was used, which comprises 25R epoxy and XN-90 carbon fiber with a high thermal conductivity of 500~W/mK, with resin weight fractions of 20\%~\cite{NT91500-520S}. The thickness of 0.10 mm with density of 2.19~g/cm$^3$ prepreg provides satisfactory mechanical strength with tensile module and strength of 860~GPa and 3430~MPa, respectively.
Table~\ref{tbl:PrepregSpecification} summarizes the specifications of the CFRP prepreg.

\begin{table}[h]
\begin{center}
\caption{Specifications of the CFRP prepreg.}
\label{tbl:PrepregSpecification}
\begin{tabular}{l|c}
\hline\hline
Item &  Specification \\
\hline
CFRP prepreg  & NT91500-520S~\cite{NT91500-520S} \\
Carbon fiber  & XN-90 \\ 
Resin & 25R epoxy \\
Resin weight fraction & 20 \% \\
Density & 2.19 g/cm$^3$ \\
Thermal conductivity & 500 W/mK \\
Tensile module & 860 GPa \\
Tensilie strength & 3430 MPa \\
Thickness & 100 $\mu$m \\
\hline\hline
\end{tabular}
\end{center}
\end{table}

\begin{figure}[htb]
\begin{center}
\includegraphics[scale=0.40]{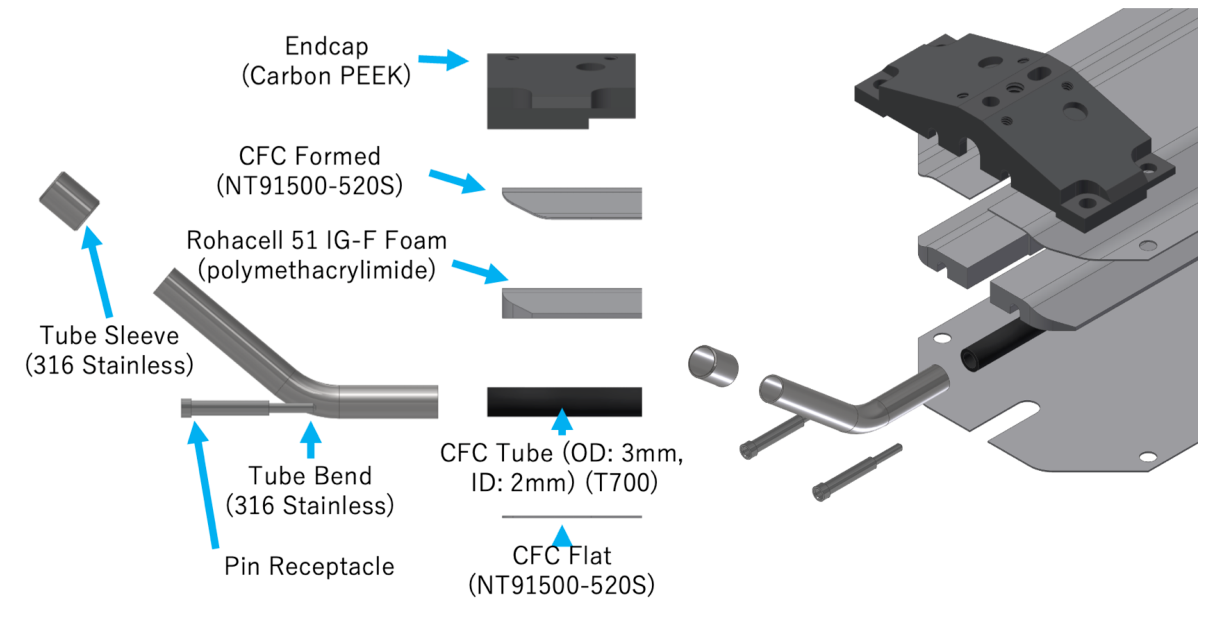}
\caption{
The mechanical design of the edge of the stave. The HDI is assembled on the bottom flat CFC plate.
\label{fig:StaveEnd}
}
\end{center}
\end{figure}

Figure~\ref{fig:StaveEnd} shows the stack up of the stave and the mechanical attachments of its edge. The shells of the stave are CFC plates at the top and bottom are flat and formed, respsectively. The formed plate is U-shaped to accommodate the 3~mm diameter cooling tube between two CFC plates with 3~mm thick ROHACELL 110 RIST form~\cite{Rohacell} that tapers down near the edge forming a structural core. The cooling tube is made of Toreyca T700C carbon fiber prepreg~\cite{T700C} fabricated by Kimuraya Co., Ltd., Japan.  
The NT91500-520S sheet is unidirectional with high thermal conductivity.  To provide omnidirectional rigidity, each CFC plate has three unidirectional sheet layers with high thermal conductivity oriented in longitudinal, transverse, and longitudinal directions. The thickness of the CFC plate is 0.33~mm and the total thickness of the stave is 3.76~mm. 

\begin{figure}[htb]
\begin{center}
\includegraphics[scale=0.40]{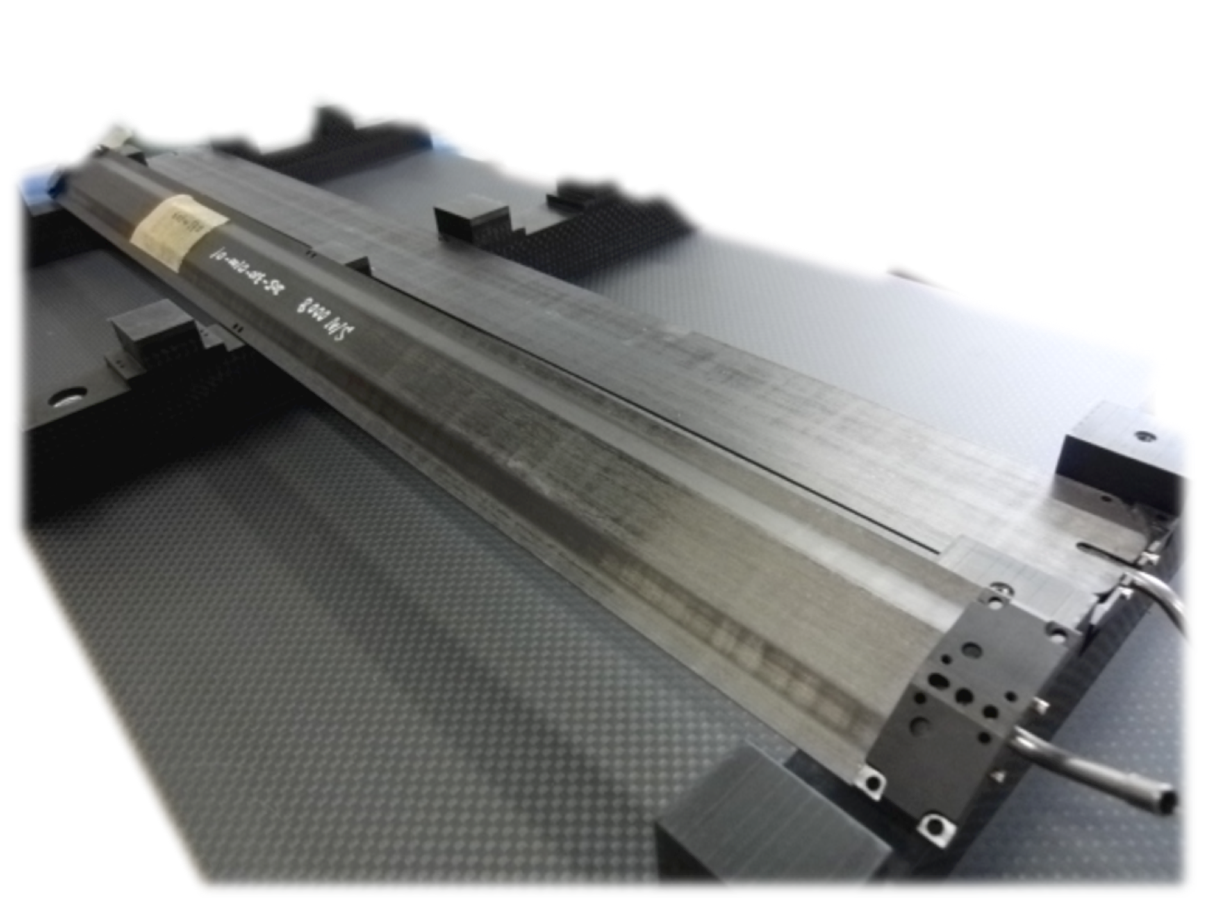}
\caption{
The photo of the INTT staves.
\label{fig:INTT_Stave_photo}
}
\end{center}
\end{figure}

\begin{figure}[htb]
\begin{center}
\includegraphics[scale=0.40]{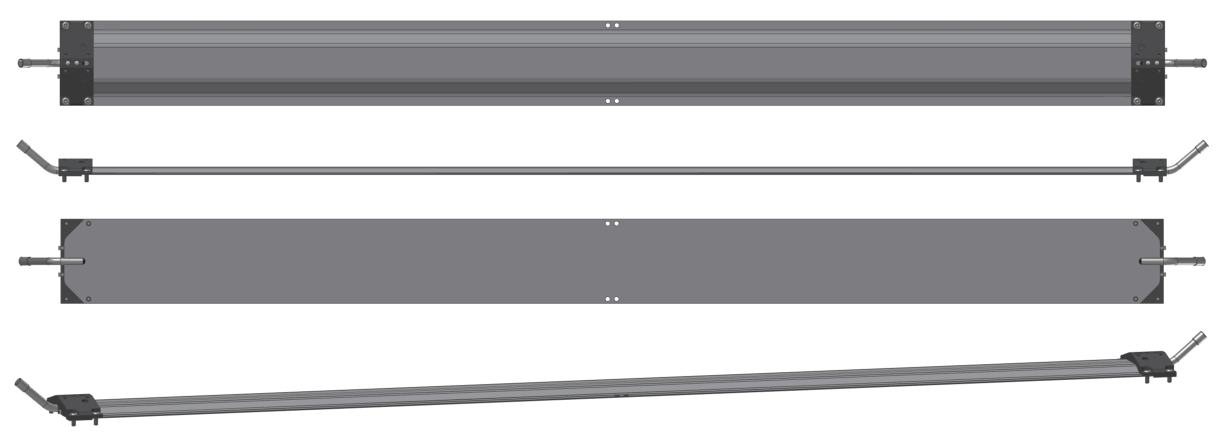}
\caption{
The INTT stave drawings from four different view angles. The HDI is assembled on the bottom flat CFC plate.
\label{fig:Stave}
}
\end{center}
\end{figure}

The mechanical attachment of both edges of the stave comprise the end caps, SUS316 cooling-tube extensions, and pairs of pin receptacles manufactured by Mill-Max MFG Co. for grounding. The endcap is made of Ketron CA30 PEEK~\cite{PEEK} manufactured by Mitsubishi Chemical Co.  The stainless tubes and the cooling tubes were glued using Henkel LOCTITE EA 9396 adhesive~\cite{EA9396}.
The rest of the stave pieces are assembled using the graphite conductive EP75-1 epoxy adhesive (MasterBond Co.)~\cite{EP75-1}.  Figure~\ref{fig:INTT_Stave_photo} and Figure~\ref{fig:Stave} show the photo and the engineering drawing of the INTT stave. Table~\ref{tbl:StaveSpecification} summarizes the material list of stave components and specifications.

\begin{table}[h]
\begin{center}
\caption{The material list of stave components and specifications. The total thickness is at the sensor region.}
\label{tbl:StaveSpecification}
\begin{tabular}{l|c}
\hline\hline
Item &  Model/Specification \\
\hline
Dimensions  & 497 mm $\times$ 38 mm \\
Total thickness & 3.76 mm \\
Adhesive for stave assembly & MasterBond EP75-1~\cite{EP75-1}\\
Structure core & ROHACELL 110 RIST~\cite{Rohacell} \\
End cap & Ketron CA3 PEEK~\cite{PEEK} \\
CFC cooling tube & Toreyca T700C~\cite{T700C} \\
Cooling tube extension & SUS316 \\
Cooling tube adhesive & Henkel LOCTITE 9396~\cite{EA9396}\\
\hline\hline
\end{tabular}
\end{center}
\end{table}

The staves were fully fabricated by Asuka Co., Ltd., Japan, including the baking process of the CFC plates.  To ensure meeting the requirements, the following four tests were applied to staves and to every cooling tube (after the stainless extensions were glued).  Only components that passed examination were used. 

\begin{enumerate}
 \item{Burst test : Tube does not burst for 1 hour at $60\pm2$~psi}
 \item{Leak test : less than 0.2 ml-mbar/min}
 \item{Heat-cycle test : $40 \leftrightarrow 0^{\circ}$C (one cycle)}
 \item{Flatness $<$ 100~$\mu$m of the flat side of the stave and alignment positions are within specified tolerance}
\end{enumerate}

 The staves were assembled with only those cooling tubes that passed the 1 and 2 examinations to keep the yield rate of the stave production reasonably high. The assembled staves were then examined for the above items 3 and 4. 
 
%%%%%%%%%%%%%%%%%%%%%%%%%%%%%%%%%%%%%%%%%%%%%%%%%%%%%%%%%%%%%%%
\subsection{Material Budget}
%%%%%%%%%%%%%%%%%%%%%%%%%%%%%%%%%%%%%%%%%%%%%%%%%%%%%%%%%%%%%%%
\label{MaterialBudget}

Table~\ref{tbl:MaterialBudget} summarizes the material budget of the silicon ladder. The total thickness of the silicon pad area of the ladder is 4.57~mm and its effective radiation length $X/X_0$ is 1.14\%. The largest contributions to the budget is the HDI (with radiation length is $X/X_0=0.43$\%), and the silicon and staves with radiation lengths $X/X_0=0.34$\% and $X/X_0=0.33$\%, respectively.

\begin{table}[h]
\begin{center}
\caption{The material budget of the silicon ladder. TC-2810 is the thermally conductive glue. The radiation thickness of the HDI listed in the table is for production batches 1 and 2.}
\label{tbl:MaterialBudget}
\begin{tabular}{l|c|c}
\hline\hline
\multirow{2}{*}{Item} &  Thickness   & Radiation Length \\
                      &       mm     &       $X/X_0$ \%  \\
\hline
Silicon Sensor & 0.32 & 0.34\\
Silver epoxy & 0.01 & 0.02\\
HDI & 0.42 & 0.43\\
TC-2810 & 0.05 & 0.02 \\
Stave & 3.76 & 0.33 \\
\hline
Total & 4.56 & 1.14 \\
\hline\hline
\end{tabular}
\end{center}
\end{table}

The material budget of the HDI is largely governed by the copper layers. As shown in  Fig.~\ref{fig:INTT_HDI_LayerCrossSection}, there are 7 layers of 9~$\mu$m thick copper which total 63~$\mu$m.  Furthermore,  both top and bottom surface layers are plated with 15~$\mu$m thick copper\footnote{The thickness of the copper plate was 20~$\mu$m only for the last production batch, which increases the radiation length of the third batch HDIs by $X/X_0=0.02$\%.}. While the simple stack up of these copper sheets results in 93~$\mu$m, the effective amount of copper in the signal layers is much less than solid ground layers. 

Table~\ref{tbl:EffectiveCuThickness} summarizes the effective copper thickness of each layer, which was estimated based on the residual copper fraction after the etching process of the HDI fabrication. The effective total thickness is estimated to be 44.7~$\mu$m, approximately 48\% of the 93~$\mu$m total stack-up thickness, with corresponding radiation length $X/X_0$ is 0.31\%. The total thickness of polymide and glue layers is 325~$\mu$m with $X/X_0$ of 0.11. The total radiation length of the HDI in the silicon pad area is thus 0.43\%.

\begin{table}[h]
\begin{center}
\caption{The effective thickness of the HDI's copper layers.}
\label{tbl:EffectiveCuThickness}
\begin{tabular}{l|c|c|c}
\hline\hline
\multirow{2}{*}{Layer} & \multirow{2}{*}{Function} & Residual copper & Effective   \\
      &          &   fraction \%   & thickness $\mu$m \\
\hline
1 & Bias + AGND & 71.3 & 17.1 \\
2 & AGND        & 93.5 & 8.42 \\
3 & Signal      &  6.5 & 0.59 \\
4 & Power       & 94.0 & 8.64 \\
5 & Signal      &  7.2 & 0.65 \\
6 & DGND        & 93.2 & 8.38 \\
7 & Bias+Signal &  4.7 & 1.12 \\
\hline
\multicolumn{3}{c}{Total}   & 44.7 \\
\hline\hline
\end{tabular}
\end{center}
\end{table}

The silicon sensors were glued on the bias pads of the top layer of the HDI using electrically conductive silver epoxy, i.e. Henkel LOCTITE ABLESTIC 2902 adhesive. The average thickness of the epoxy was designed to be 9~$\mu$m based on the 50~$\mu$m thick glue mask, which has total of 295 glue potting holes with 2 mm diameter each.
The radiation length of the silver epoxy was then calculated to be $X/X_0=0.02$\% based on the mixing ratio of the silver powder and the epoxy of 0.21:0.79~\cite{SilverEpoxy}. Due to the short radiation length $X_0=8.54$ mm of the silver material, the contribution of the silver epoxy to the total material budget can be non-negligible. Without the glue mask, the contribution of the 50~$\mu$m thick silver epoxy is as significant as 12\% of the total material budget.

To diffuse the heat generated by the FPHX chips, highly thermally conductive epoxy adhesive glue, 3M$^{\rm TM}$ TC-2810~\cite{TC-2810}, 
was used to assemble the HDI and the stave.  The radiation length for the 50~$\mu$m thick the epoxy makes a very small contribution to the total material budget of the silicon ladder, unlike the silver compound one.

%%%%%%%%%%%%%%%%%%%%%%%%%%%%%%%%%%%%%%%%%%%%%%%%%%%%%%%%%%%%%%%
\subsection{Dead Space}
%%%%%%%%%%%%%%%%%%%%%%%%%%%%%%%%%%%%%%%%%%%%%%%%%%%%%%%%%%%%%%%
\label{Dead Space}

\begin{figure}[htb]
\begin{center}
\includegraphics[scale=0.40]{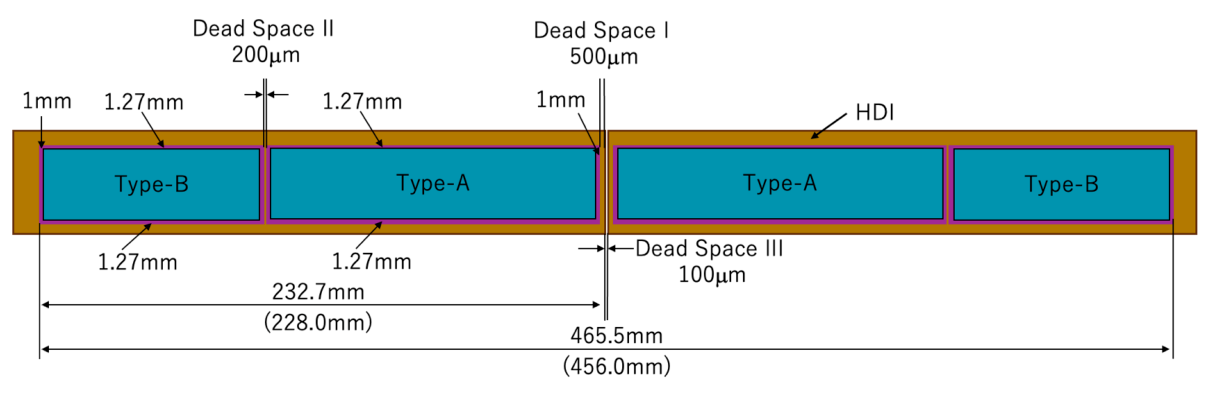}
\caption{
Dead spaces of the INTT ladder. The number in parentheses is the longitudinal length of the active region of the type-A and B sensors altogether.   
\label{fig:INTT_Ladder_DeadSpace}
}
\end{center}
\end{figure}

Due to the precision limit of the silicon sensor alignment during the assembly process of the silicon sensors on the sensor pads of the HDI, there are small dead spaces between the edge of the type-A sensor and the tip of the HDI (dead space I), and also the type-A and type-B sensors (dead space II). They are 0.5 and 0.2 mm, respectively, as shown in Figure~\ref{fig:INTT_Ladder_DeadSpace}.

As the result of the dead spaces in the silicon alignment, the total distance from the boundary of the two half ladders to the large rapidity end of the type-B sensor is 232.7 mm, while the length of the active region of type-A and type-B sensors is 228.0 mm as tabulated in Table~\ref{tbl:SiliconDemensions}. This results in 98\% coverage of the active region in the longitudinal direction as summarized in Table~\ref{tbl:ActiveRegionLongitudinal}. On the contrary, there is no additional dead space originating from the alignment in the transverse direction. The fraction of the active region in the transverse direction is thus 88.7\% as tabulated in Table~\ref{tbl:ActiveRegionTransverse}. The resulting fraction of the active area of the half ladder is 87.0\% as tabulated in Table~\ref{tbl:ActiveRegionHalfLadder}.

\begin{table}[h]
\begin{center}
\caption{The dimensions of the dead space and region in the longitudinal direction of the half ladder.}
\label{tbl:ActiveRegionLongitudinal}
\begin{tabular}{l|c|c}
\hline\hline
Longitudinal              & Physical & Active \\
\hline
Dead space I              & 0.5 mm & 0 mm\\
Type-A                    & 130 mm & 128 mm\\
Dead space II             & 0.2 mm & 0 mm \\
Type-B                    & 102 mm & 100 mm \\
Total                     & 132.7 mm & 228.0 mm\\
\hline
\multicolumn{2}{c}{The fraction of longitudinal active length} & 98.0\% \\
\hline\hline
\end{tabular}
\end{center}
\end{table}

\begin{table}[h]
\begin{center}
\caption{The physical and active area of the half ladder.}
\label{tbl:ActiveRegionTransverse}
\begin{tabular}{l|c|c}
\hline\hline
Transverse                & Physical & Active \\
\hline
Type-A \& B               & 22.5 mm & 19.968 mm\\
\hline
\multicolumn{2}{c}{The fraction of transverse active length} & 88.7\% \\
\hline\hline
\end{tabular}
\end{center}
\end{table}

\begin{table}[h]
\begin{center}
\caption{The physical and active region of the half ladder.}
\label{tbl:ActiveRegionHalfLadder}
\begin{tabular}{l|c|c}
\hline\hline
                & Physical & Active \\
\hline
Area            &  5235.75 mm$^2$ & 4552.70 mm$^2$ \\
\hline
\multicolumn{2}{c}{The fraction of active area} & 87.0\% \\
\hline\hline
\end{tabular}
\end{center}
\end{table}

Lastly, there is another 0.1 mm of dead space between two half silicon ladders (Dead Space III in Figure~\ref{fig:INTT_Ladder_DeadSpace}) originating from the alignment precision of the two half silicon modules on the stave. The resulting physical length between the large rapidity end of the type-B sensors of two half ladders is 465.5 mm, while the total longitudinal length of the active regions of 4 silicon sensors is 456.0 mm. This dead space is relatively minor enough, the fraction of active area still stays as 98.0\% in the longitudinal direction for the full ladder. 

%%%%%%%%%%%%%%%%%%%%%%%%%%%%%%%%%%%%%%%%%%%%%%%%%%
%%%%%%%%%%%%%%%% INTT Readout %%%%%%%%%%%%%%%%%%%%
%%%%%%%%%%%%%%%%%%%%%%%%%%%%%%%%%%%%%%%%%%%%%%%%%%
\section{INTT Readout Cables and Front End Circuit Board}
\label{INTT Readout}

Because the entire INTT barrel must fit within the inner diameter of the TPC detector~\cite{Aidala:2012nz}, the signal from the barrel must be transmitted all the way to the outside of the TPC volume.  The ROCs from the FVTX detector, which are downstream electronics for signal processing, cannot fit within the inner diameter of the TPC. 

Massive raw data generated from the INTT must be transmitted at high-speed to the ROC through a curved cable path for longer than 1 m.  Because no commercial cable satisfies the requirements, a novel bus-extender cable was developed based on the FPC. This technology simultaneously satisfies the requirements of high-density signal lines, flexibility, and long cable length.  In contrast, the specification of line and space of the bus extender prevented its connector-end design to be compatible with the input connector ports of the existing ROC. Therefore, another 15 to 25~cm conversion cable was developed to interconnect between the bus extender and the ROC. Flexibility in three-dimensions for the conversion cable is required to guide the connection smoothly without introducing any stress at the ROC input connector. This is needed to absorb the geometrical mismatch of the downstream end of the bus extender and the corresponding input connector location of the ROC. 

\begin{figure*}[htb]
\begin{center}
\includegraphics[scale=0.80]{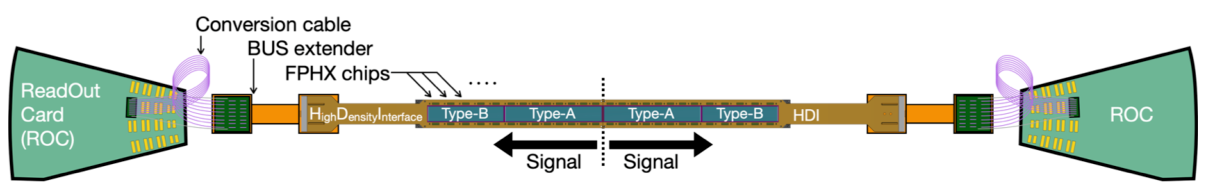}
\caption{
The schematics of the INTT silicon ladder and full readout cable chain up to the ROC.
\label{fig:INTT_Ladder_ReadoutChain}
}
\end{center}
\end{figure*}

The schematics of the INTT silicon ladder and full readout cable chain up to the ROC is depicted in Fig.~\ref{fig:INTT_Ladder_ReadoutChain}. The data of each half ladder is readout from opposite side of the full ladder and processed by different ROCs.  

%%%%%%%%%%%%%%%%%%%%%%%%%%%%%%%%%%%%%%%%%%%%%%%%%%%%%%%%%%%%%%
\subsection{Bus Extender Cable}
%%%%%%%%%%%%%%%%%%%%%%%%%%%%%%%%%%%%%%%%%%%%%%%%%%%%%%%%%%%%%%
\label{BEX}
 
\begin{figure}[htb]
\begin{center}
\includegraphics[scale=0.40]{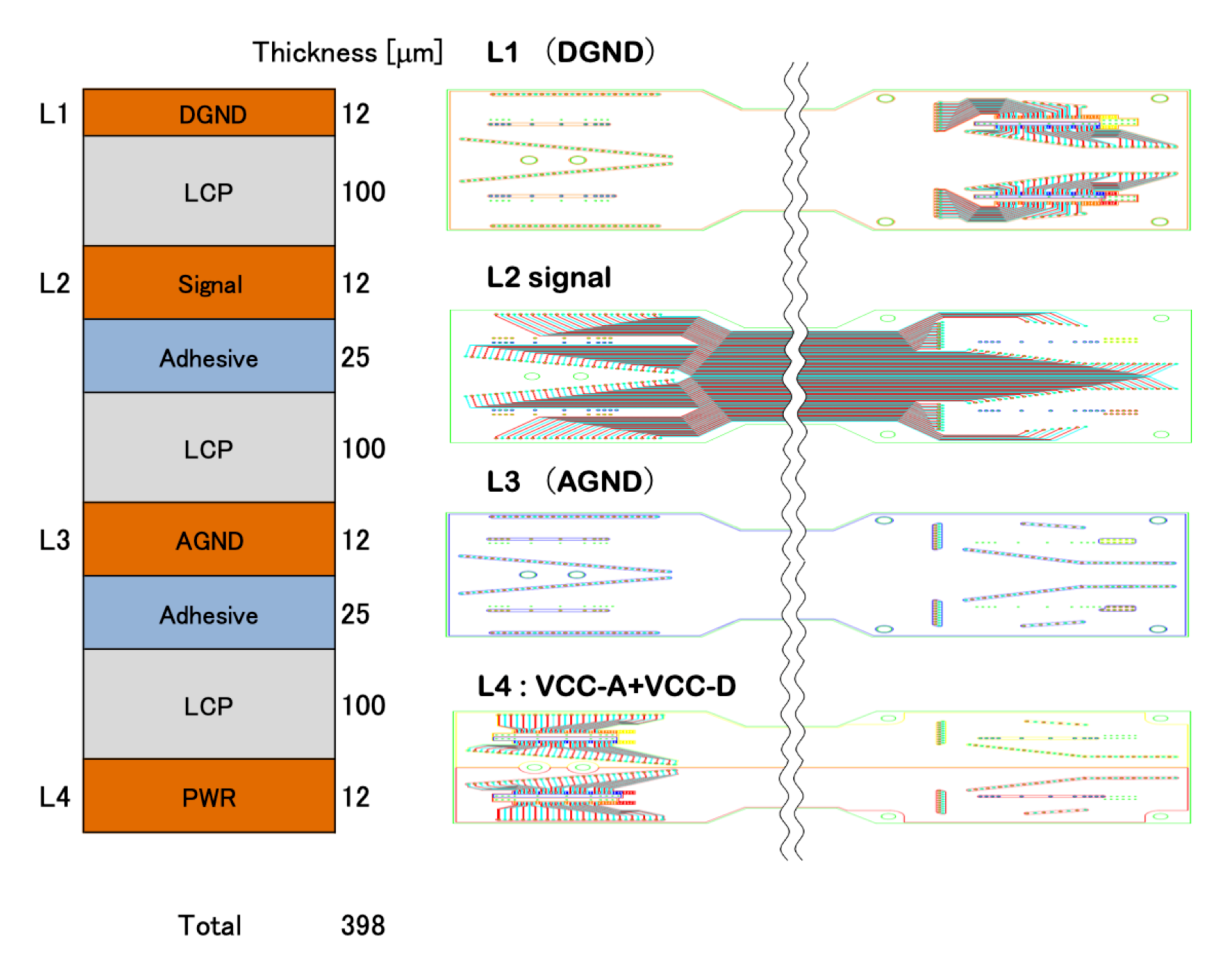}
\caption{
Four layer structure of the BEX (left), and schematics of each conductive layer (right) from L1 to L4.
\label{fig:INTT_BEX_Layers}
}
\end{center}
\end{figure}

Only key features of the bus extender (BEX) are described here, because details of the BEX cable are discussed  elsewhere~\cite{BEX}. 
This FPC cable comprises four layers of three flexible-copper-clad laminate (FCCL) as shown in Fig.~\ref{fig:INTT_BEX_Layers}. The top and third layers are for the digital and analog grounds, respectively. The second layer is used for signal lines.  The bottom layer has the power supply lines for the FPHX chips.
The choice of the four-layer structure, instead of the seven-layer structure of the HDI, is primarily driven by fabrication constraints. 
The yield rate is a big concern for the BEX. The main driver is the signal layer due to its required microfabrication accuracy over the extraordinary length of 1.11 meters. Reducing the number of signal layers was chosen to minimize the risk in the fabrication process.  All of the BEX signal lines were thus integrated into a single layer. 

Polymide is the most popular market choice of a dielectric material for the FPCs and its fabrication technology is well established in industry.  For the BEX case, liquid crystal polymer (LCP) was chosen as the dielectric material due to its lower transmission loss of signal amplitude and the availability of thicker FCCLs in the industrial market compared to polymide.   These features are advantageous for the bus extender, whose design is strictly constrained by the precision and yield limit in the fabrication process due to its extraordinarily long length.  

The realistic limit of the line and space for the BEX were both found to be 130~$\mu$m.  To achieve a reasonable yield rate in the fabrication factory, which was Printed Denshi Kenkyusho Co., Ltd.   Given the line and space, the thickness of the dielectric material is required to be as thick as 100~$\mu$m to match the differential characteristic impedance of 100~${\rm \Omega}$. While such a thick polymide is not available in the industrial market, the 100~$\mu$m thickness is available in the FELIOS model LCP from Panasonic Industry Co., Ltd.~\cite{FeliosLCP}. The key specifications of the LCP to make the low transmission loss possible are the dielectric constant $\epsilon_{\rm r}=3.3$ and the dissipation factor $\tan\delta=0.002$. 

For the bonding sheet, a specially optimized model for LCP was employed 
%model A26R of Arisawa Manufacturing Co., ltd.~\cite{Arisawa} 
with thickness of 25~$\mu$m chosen for having a low dielectric constant. As a consequence of R\&D, this was also a crucial choice to achieve decent yield rate in the through-holes plating process~\cite{BEX} by Taiyo Manufacturing Co., Ltd. because the fabrication technology for the LCP is not as well established as for the polymide. Table~\ref{tbl:BEXSpecification} summarizes the specifications of the BEX cable. 

\begin{table}[h]
\begin{center}
\caption{Specifications of the BEX cable.}
\label{tbl:BEXSpecification}
\begin{tabular}{l|c}
\hline\hline
Item &  Specification \\
\hline
Dimension & 1.11 m $\times$ 34 mm \\
Width (connector region) & 43 mm \\
Total thickness & 398 $\rm \mu$m \\
\hline
Dielectric material & LCP\\
Dielectric model & Panasonic FELIOS~\cite{FeliosLCP}\\
Dielectric layer thickness & 100 $\mu$m \\
%\hline
%Bonding sheet model & A26R~\cite{Arisawa}\\
%Bonding sheet make & Arisawa Manufacturing \\
Bonding sheet thickness & 25 $\mu$m \\
%\hline
Number of layers & 4 \\
Number of signal lines & 122 \\
Line and space & 130 \& 130 $\mu$m \\
\hline\hline
\end{tabular}
\end{center}
\end{table}

\begin{figure}[h]
\begin{center}
\includegraphics[scale=0.9]{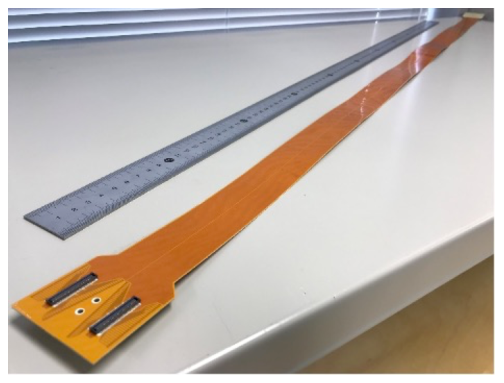}
\caption{
A photograph of the 1.11 meter long BEX cable. 
\label{fig:INTT_BEX_photo}
}
\end{center}
\end{figure}

Pairs of DF18C-100DS-0.4V(81) receptacle connectors manufactured by Hirose Electric Co. Ltd. are implemented at  both ends of the bus extender as shown in Fig.~\ref{fig:INTT_BEX_photo}. The space between a pair of DF18 plugs is 26~mm which is incompatible with the 10 mm~spacing of the input connector pair of the ROC. This extra spacing between the connector pairs of the BEX is caused by the difficulty to wire 122 signal lines of 130~$\mu$m line~\&~space into the short connector spacing. Due to the incompatibility of the connector layout between the BEX and the ROC, a conversion cable was introduced which converts the connector spacing from the BEX to the ROC spacing. 

\begin{figure}[htb]
\begin{center}
\includegraphics[scale=0.9]{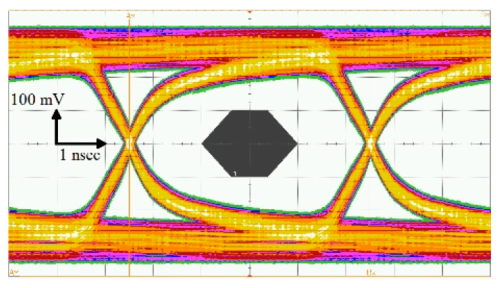}
\caption{
The measured eye-diagram of the BEX as a result of transmitting 200~MHz signals for 1 million times~\cite{BEX}.   
\label{fig:INTT_BEX_EyeDiagram}
}
\end{center}
\end{figure}

The bus extender successfully moderated the signal attenuation. However, the signal level is still questionable due to the extraordinarily long cable length as a multilayered FPC. The performance of signal transmission was evaluating with measurements using each of the following methods: 1) S-parameters, 2) eye-diagram, and 3) time-domain reflectometry (TDR). From the S-parameter measurement, the insertion and reflection losses are -2.7 and -23~dB, respectively at 200 MHz. The measured eye diagram is shown in Fig.~\ref{fig:INTT_BEX_EyeDiagram}. The waveform is confirmed to exhibit a sufficient margin to a defined mask\footnote{Private communication with the FPHX chip developers.} by observing 1 million waveforms of the 200 MHz signal (see the solid hexagon in the middle of Fig.~\ref{fig:INTT_BEX_EyeDiagram}).
The measured characteristic impedance was 90~${\rm \Omega}$ differential in the TDR measurement. This is 10\% smaller than the default 100 ${\rm \Omega}$, it is confirmed to be permissible by the return loss measurement of the daisy chain with the conversion cable as discussed in subsection~\ref{Conversion Cable}.  Table~\ref{tbl:BEXPerformance} summarizes the performance of the BEX cable.

\begin{table}[h]
\begin{center}
\caption{The performance of the BEX cable. The insertion loss and return losses are the performance at 200~MHz. The characteristic impedance is the differential of the LVDS pair.}
\label{tbl:BEXPerformance}
\begin{tabular}{l|c}
\hline\hline
Item &  Performance \\
\hline
Insertion Loss  & -2.7 dB\\
Return Loss & -23 dB \\
Characteristic impedance  & 90 ${\rm \Omega}$\\
\hline\hline
\end{tabular}
\end{center}
\end{table}

%%%%%%%%%%%%%%%%%%%%%%%%%%%%%%%%%%%%%%%%%%%%%%%%%%%%%%%%%%%%%%
\subsection{Conversion Cable}
%%%%%%%%%%%%%%%%%%%%%%%%%%%%%%%%%%%%%%%%%%%%%%%%%%%%%%%%%%%%%%
\label{Conversion Cable}

As discussed in previous sections, the FPC is the first choice of technology to satisfy the high-performance requirements of signal transmission, high signal-line density, and flexibility. 
However, it is difficult for the FPC to satisfy an additional and unique requirement as the last stage of the readout cable series.  The requirement is flexibility in three dimensions to connect the downstream end of the BEX and the input connector ports of the ROC board without introducing any stress at the connection. 
Due to the geometrical mismatch between the INTT barrel ladders and the input connector layout of the ROC, flexibility in three dimensions is crucial for the conversion cable. 
If the FPC cable has directivity, then stress is induced at the connector end, which would bend the cable to the transverse direction of the cable plane.  

A $\mu$-coax technology was chosen. Although the signal line density cannot be as high as the FPC, the three dimensional flexibility offers a good trade-off in the INTT readout design. According to an engineering study, due to its lack of flexibility the FPC solution would require as many as 14 different designs in curving and length to interconnect mismatching connector geometries between the BEX and input connectors of the ROC.  Furthermore, spare cables of all designs would be needed for risk management, which leads to high cost inefficiency. The $\mu$-coax solution reduces the number of designs to be only two in different lengths.  

The CABLINE-UX II model manufactured by I-PEX Inc.~\cite{IPEX} was employed as the last stage of the signal transmission cable chain for the INTT ladder. The AWG$\#44$ harness %manufactured by Aosen Co. 
is made of silver-plated copper alloy with a center conductor insulated by 30$\mu$m thick perfluoroalkoxy alkane (PFA) dielectric material from tinned copper alloy with wire wrapping in the outer spiral shield. The shield is covered by the outermost jacket made of the PFA, as well forming a four-layered coaxial cable with total thickness of $0.24\pm0.01$~mm. The characteristic impedance of the harness is 45~$\Omega$ (90 differential for a LVDS pair). 
A slim plug and small connectors are assembled for both ends of the harnesses to bundle 50 harnesses in the wire spacing with 0.25~mm pitch. Finally, the harness bundle is wrapped by an acetate cloth adhesive tape. 

\begin{figure}[h]
	\centering
		\includegraphics[scale=0.4]{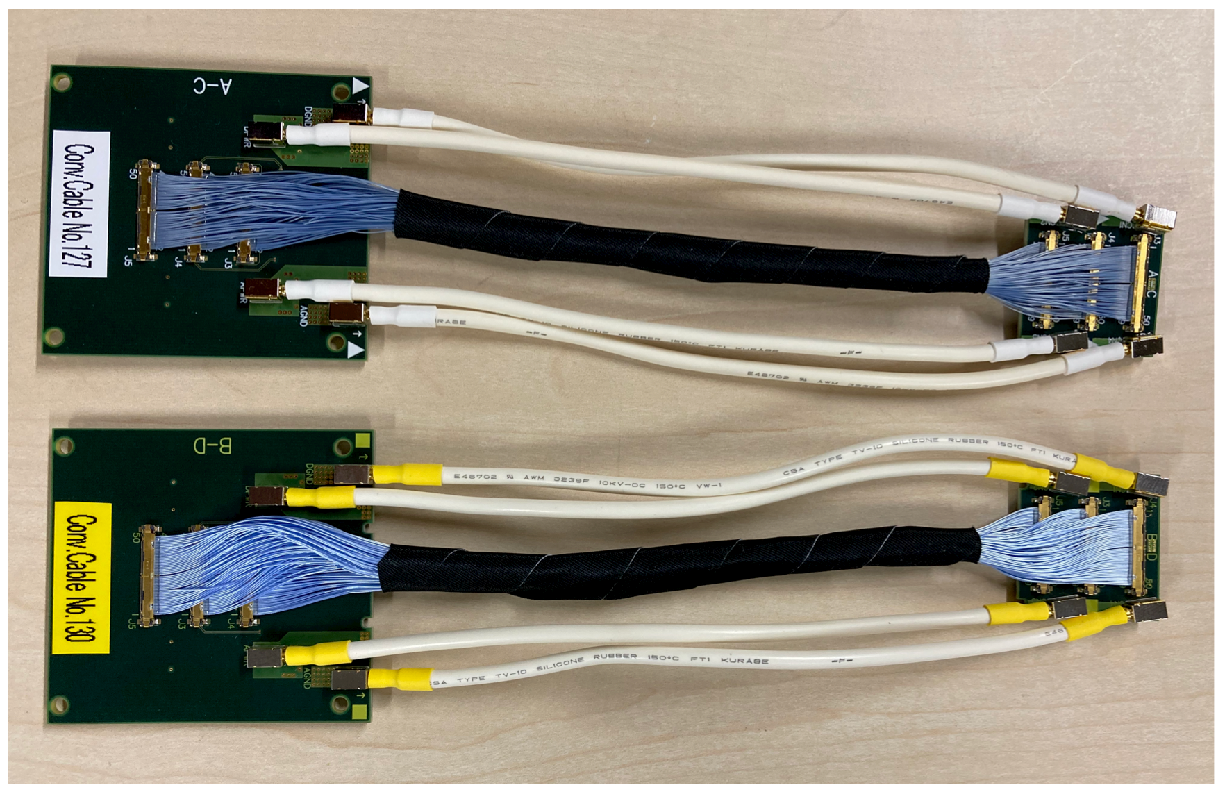}
	  \caption{Conversion cable type-AC (top) type-BD (bottom). The bundles in the middle of the cable are 15~cm length $\mu$-coax harnesses with four white jacket cables for the power and ground cables. The larger PC board is the connector for the BEX side and the smaller board is for the ROC side.}
      \label{ConversionCable}
\end{figure}

The conversion cable interconnecting between the BEX and ROC comprises three $\mu$-coax bundles, two power and two ground cables, and the PC boards in both ends. The AWG$\#24$ power and ground cables in both ends are assembled with HJ-3 male pins manufactured by MAC EIGHT Co. Ltd.. The male pin and the receptacle connection are secured by the HH-3-R lock. The wire gauge is optimized to be AWG$\#24$ to drop the voltage properly and to provide the FPHX chip power from the slightly higher regulator voltage output at the ROC. Details of the voltage drop is discussed in Section~\ref{ROC}. 

Three UX II receptacles and four HH-3-G sockets (manufactured by MAC EIGHT Co. Ltd.) are implemented on the top side of the PC board and two DF18C-100DS-0.4V(81) receptacle connectors are implemented on the bottom side of the board. The PC board is fabricated by Hayashi-REPIC Co. Ltd. and eight layered with dedicated layers for the digital and the analog-power and ground. The dimensions are 48~mm wide~$\times$~52~mm long for the BEX end to match the size of the connector end, while the ROC side is rather compact with dimensions 25~mm wide~$\times$~25~mm long. There are two types of conversion cables namely "type-AC" and "type-BC" and the difference is so the channel mapping is compatible with the channel map, which are differently designed for the columns~A$\&$C and B$\&$D of the input ports on the ROC.  There are two different lengths of 
harnesses for the power and ground cables, which are 15 and 25~cm, respectively.

\begin{table}[h]
\begin{center}
\caption{Specifications of the conversion cable. The characteristic impedance is given in differential of the LVDS pair.}
\label{tbl:CCSpecification}
\begin{tabular}{l|c}
\hline\hline
Item &  Specification \\
\hline
$\mu$-coax make & I-PEX Inc. \\
$\mu$-coax model & CABLINE-UX II~\cite{IPEX} \\
%Harness model & Aosen co. SP-10005-027A \\
Harness AWG & 44 \\
Characteristic impedance & 90  ${\rm \Omega}$  \\
Number of harness per bundle & 50 \\
Number of bundle & 3 \\
Power/ground cable AWG & 24 \\
Number of power cable & 2 \\
Number of ground cable & 2 \\
Length  & 15 \& 25 cm \\
\hline\hline
\end{tabular}
\end{center}
\end{table}

%%%%%%%%%%%%%%%%%%%%%%%%%%%%%%%%%%%%%%%%%%%%%%%%%%
%%%%%%%%%%%%%% INTT Electronics %%%%%%%%%%%%%%%%%%
%%%%%%%%%%%%%%%%%%%%%%%%%%%%%%%%%%%%%%%%%%%%%%%%%%
%%%%%%%%%%%%%%%%%%%%%%%%%%%%%%%%%%%%%%%%%%%%%%%%%%%%%%%%%%%%%%
\subsection{ROC}
%%%%%%%%%%%%%%%%%%%%%%%%%%%%%%%%%%%%%%%%%%%%%%%%%%%%%%%%%%%%%%
\label{ROC}

The readout card (ROC) is a multilayered circuit board implemented outside of the TPC volume. The ROCs were refurbished after the FVTX operation in the PHENIX experiment and reused for the INTT. ROC details are described elsewhere~\cite{Aidala:2013vna}. Any aspects of ROC usage that are different from the FVTX or customized for sPHENIX are discussed below. 

Twelve pairs of DF18C-100DP-0.4V(51) plug connectors and four pairs of DF18C-60DP-0.4V(51) are implemented in the ROC as input ports for a half ladder. The DF18C-60DP-0.4V(51) plug connector has 60 pins and is not used for the INTT. 

The ROC boards are originally implemented with 2.5 V regulators to provide analog and digital power for the FPHX chips with an operating voltage of 2.5 V for both analog and digital~\cite{PHENIX_FPHX_cite, Kapustinsky:2010nim, FPHX_Manual}. The voltage drop in the power-transfer line was minor in the FVTX owing to the short readout-cable length and low power consumption of the FPHX chip.  On the contrary, as has been discussed in the previous sections, the INTT operates in the higher-power consumption mode due to its extraordinarily long readout cable chain.  The FPHX operates with higher LVDS current for output data transmission than do the FVTX chips so that the signal amplitude is kept well above the receiver driver threshold.    

The maximum drawing current per half ladder (26 FPHX chips) for the analog and digital power of the FPHX chips is measured to be 0.21 and 0.42~amps, respectively. Although the ROC are designed to allocate reasonably large physical cross sections or american wire gauge (AWG) for the power transmission lines, the resistances of these cables is a few hundred m$\rm \Omega$.  The resulting voltage drop of the daisy chain of the readout cables amounts to as much as 0.2 to 0.4~V which is sizable enough for the FPHX chips to possibly malfunction due to insufficient voltage supply.

%The resistance and voltage drop $\Delta V$ for each cables are summarized in the Table~\ref{tbl:VoltageDrop}. The resistances of the BEX and CC were measured using the 4-terminal sensing method. The resistance of the HDI was calculated based on the cross section area of the power layers of HDI and the electric resistivity of the copper (1.68 ${\rm \Omega}$m)~\cite{CopperResitance} for the length of 39.8 mm. The length of the power line depends on the longitudinal location of the FPHX chip on the HDI. The effective resistance to the closest FPHX to the connector end is to be approximately half of 0.1 ${\rm \Omega}$. 
%The voltage drops for the analogue and digital in each cables are calculated based on the corresponding drawing currents, respectively. The total voltage drops are 0.20 and 0.42, respectively, which results in lower voltage at FPHX chip by 16\% and 34\% of voltage 2.5 V including the voltage drop in the return path as well. This level of voltage drop is fatal to operate FPHX chips. 
To compensate for the voltage drop, the surface mounted regulators for the FPHX power are upgraded to supply larger output voltages. The pin layouts of the upgraded regulators match the pad pattern of the ROC.  Newly implemented regulators are MCP1700-2802E/TT and MCP1726-3002E/MF (Microchip Technology Inc.) with outputs of 2.8 and 3.0~V for analog and digital power, respectively.  With drawing currents increased for higher voltage outputs, the supplied voltages at the FPHX location are calculated to be $\approx$ 2.5 to 2.6~V for both analog and digital.

%%%%%%%%%%%%%%%%%%%%%%%%%%%%%%%%%%%%%%%%%%%%%%%%%%
%%%%%%%%%%%%%%%% Ladder Performance %%%%%%%%%%%%%%
%%%%%%%%%%%%%%%%%%%%%%%%%%%%%%%%%%%%%%%%%%%%%%%%%%
\section{Radiation Hardness}
\label{RadiationHardness}

In the INTT readout system (see Section~\ref{INTT Readout}), there are a few materials for which radiation hardness is either not known or known to be not durable. In this section, the radiation hardness is evaluated for these items.  Also discussed is the study of the potential radiation damage of the ROC boards during the preceding PHENIX operation. 

\subsection{Bus Extender}
The FPC with standard polymide is widely used in radiation environments and its radiation hardness is well established. 
However, because LCP is a relatively new material being used for the dielectric layer of the FPC, the radiation hardness of the FPC with LCP has not been established as well as that of the polymide. The radiation hardness of the LCP material itself is proven to be as durable as polymide~\cite{Whinnery:2017snd} although, the radiation hardness of the bonding sheet employed for the BEX was not known. The 
overall mechanical characteristics of the BEX after assembly was measured before and after a radiation exposure rather than investigating the radiation hardness of the bonding sheet itself. The primary concerns are the degradation in the flexibility of the BEX and the peel strength of the bonding sheet due to radiation damage.

\begin{figure}[htb]
\begin{center}
\includegraphics[scale=0.45]{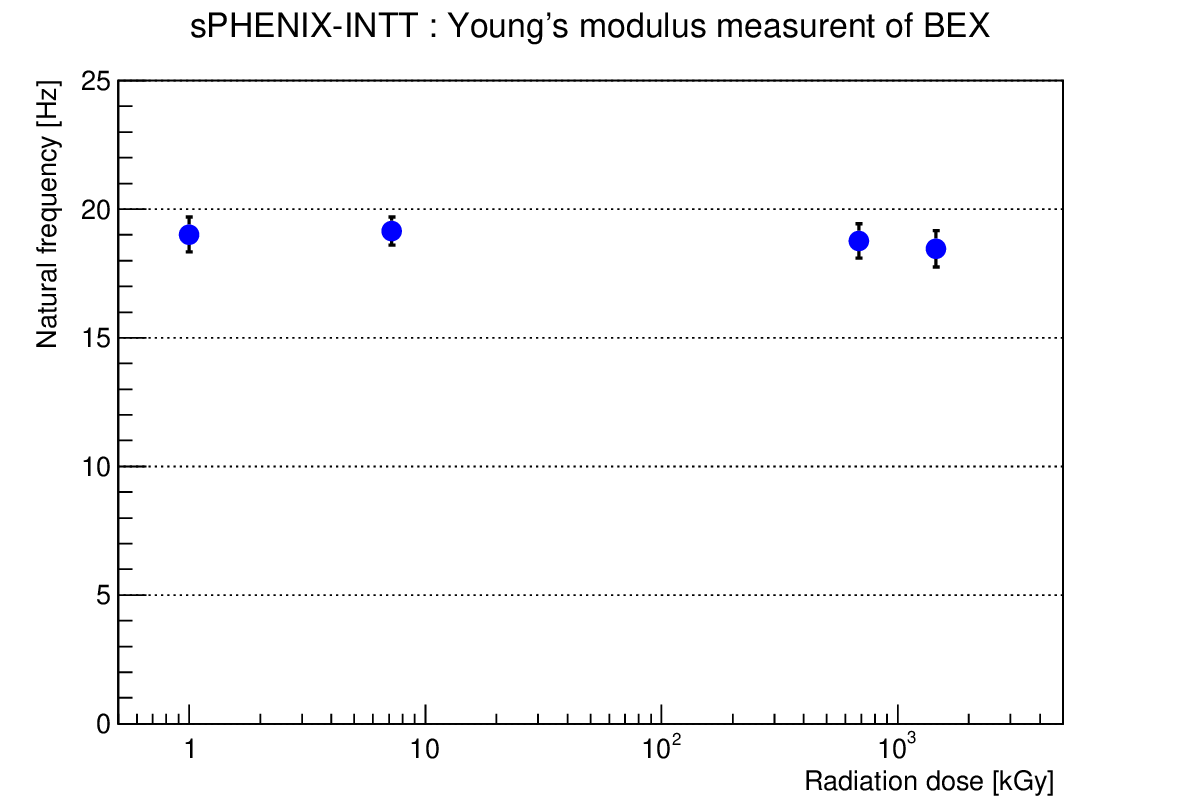}
\caption{
The Young's modulus measurement results of samples exposed to various radiation dose. The vertical axis is the natural frequency in the unit of Hz, while the horizontal axis is the radiation dose in the unit of kGy. The measurement of the sample with no radiation exposure is plotted purposely at 1 kGy.
\label{fig:YoungsModulus}
}
\end{center}
\end{figure}

A few samples of the BEX cable were exposed to $^{60}$Co source for 7.2 kGy, 685 kGy, 1.45 MGy at National Institutes for Quantum Science and Technology, Japan. The expected radiation dose at the location where the BEX is installed in sPHENIX is approximately 5 kGy for five years operation\footnote{Including an extra two years of operation beyond the officially approved three years.}. 
The mechanical performances were evaluated by the following two tests: 1) stress test to evaluate the flexibility, and 2) peel strength test between the copper and the LCP layers~\cite{BEX}. As the test item 1), the Young's modulus was measured for the samples before and after the radiation exposure. Figure~\ref{fig:YoungsModulus} shows the Young's modulus of these samples based on natural frequency measurements.  No degradation of flexibility was observed within the accuracy of the measurements (7\%) in any samples. The error was estimated by the standard deviation of multiple samples within the same radiation dose group.

\begin{figure}[htb]
\begin{center}
\includegraphics[scale=0.45]{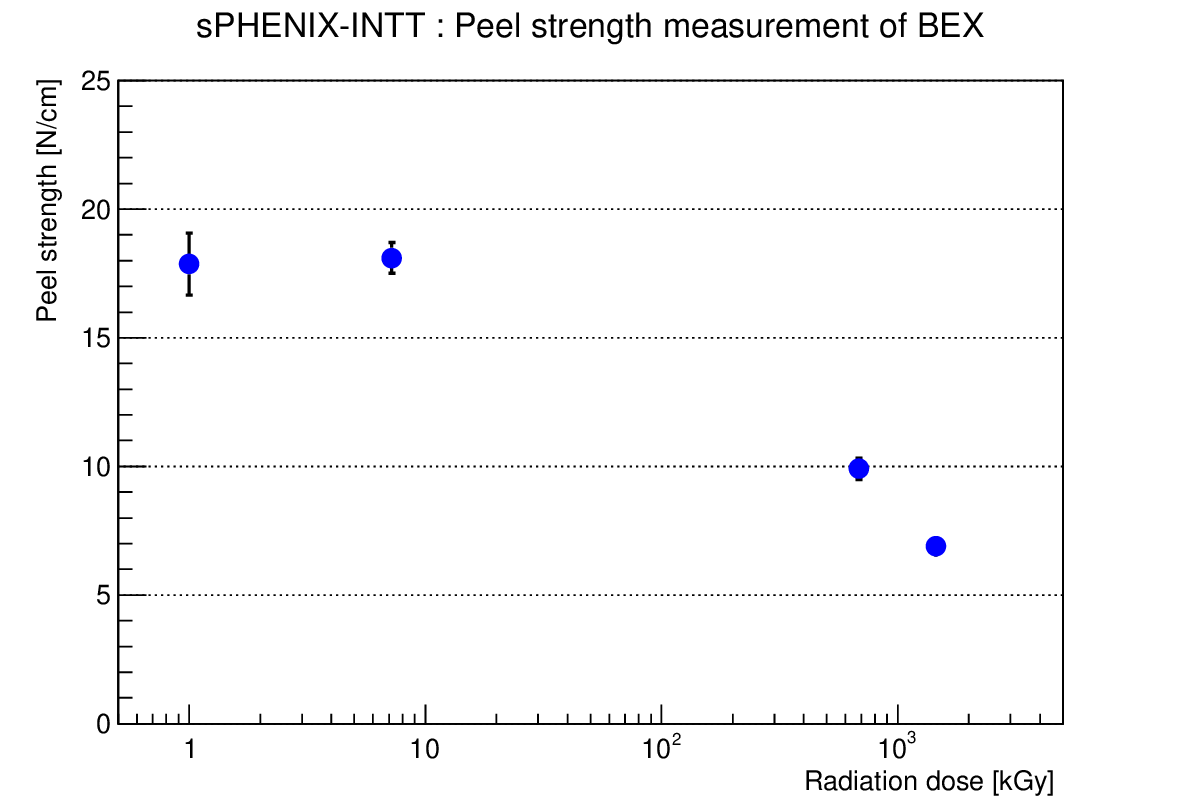}
\caption{
The peel strength measurement results of samples exposed to various radiation doses. The vertical axis is the peel strength in units of N/cm, while the horizontal axis is radiation dose in units of kGy. The measurement of the sample with no radiation exposure is plotted at 1 kGy on purpose.
\label{fig:PeelStrength}
}
\end{center}
\end{figure}

For test item 2), approximately 50\% and 70\% degradations were observed in the peel strength for 685 kGy and 1.45 MGy samples, respectively as shown in Fig.~\ref{fig:PeelStrength}. However, the measured peel strength of 18~N/cm for the 7.2 kGy sample proved to be sufficiently strong and is similar to the typical level of polymide-based FPC. Although the safety margin to keep this peel strength for possible unexpected extra radiation dose may be marginal, it is unlikely that FCCL layers would fall apart by gravity even if the peel strength were weakened by 50\%. It was thus concluded that the radiation hardness of the BEX is expected to be durable for the duration of sPHENIX operation. 

\subsection{Conversion Cable}
The material "fluorinated resin", which is used as a dielectric insulator for the $\mu$-coax cable~\cite{IPEX}, is reported to be weak against radiation~\cite{Whinnery:2017snd} compared to popular dielectric insulator materials like polyimide or LCP. The estimated equivalent neutrons at the location of conversion cables to be installed is $0.15{\times}10^{12}$ for three years of sPHENIX operation. 

In order to address the radiation hardness of the $\mu$-coax harness for three years of sPHENIX operation, the effect of radiation was studied at the RIKEN Accelerator-driven compact neutron systems facility (RANS), Japan. Three samples of harness bundle were exposed to the RANS neutron beam at energies up to 5 MeV. Each sample was exposed to 1.3, 2.6, and 4.0~$\times~10^{12}$ equivalent neutrons. These are factors of 9, 17, and 27, respectively, more than the estimated radiation dose for the INTT operation in sPHENIX. The signal transmission performances were compared before and after the irradiation to evaluate the radiation effect.  The comparisons were made for the S-parameters, the eye diagrams, and the TDR. No degradation was observed for all samples in any of these measurements,  regardless of the exposed radiation dose. Thus, it is reasonable to conclude that the radiation hardness of the conversion cable is durable for the INTT operation as well.

\subsection{ROC}
The radiation dose of the ROC boards throughout five years of operation of the FVTX detector in PHENIX is estimated to have been approximately 300 Gy, while the corresponding radiation dosage for three years of the INTT operation in sPHENIX is estimated to be approximately 50 Gy. The dose for the INTT is moderate owing to the relatively farther distance of the ROC position from the collision point compared to that of the FVTX in PHENIX. Here the radiation hardness of the ROC board with respect to the total dose of 350 Gy is evaluated. 

The ROC board is designed to be radiation hard and in fact it utilizes the FLASH-based ACTEL ProASIC3E FPGAs~\cite{Aidala:2013vna} which is known to be radiation hard. The optical data transmission system of the ROC board comprises the model TLK2711 (Texas Instruments) as the serializer/deserializer of data. There is a study of the radiation tolerance of the TLK2711~\cite{TLK2711}. In this study, the increase of leakage current and the bit-error rate was monitored as a function of radiation dose. The first bit error appeared $\approx$280--420~Gy and the TLK2711 encountered functional failure as low as 700~Gy depending on the beam condition of the radiation exposure, e.g., high (low) intensity and short (long) duration. While the leakage current stays the same up to 400 Gy, a rather rapid increase of current was observed beyond that point. 

According to the above study, the condition of the TLK2711 are already in the range that some bit error symptom may start after the FVTX use. Altough it was preferable to replace them all before the reuse for the INTT, the model was discontinued. Unfortunately, insufficient quantities were available in the market place to replace them all.  Hence, the replacement candidates are prioritized and limited to only those which already have bit-error symptoms. The only ROCs installed are those which passed the various function tests including no bit-error symptoms.

%%%%%%%%%%%%%%%%%%%%%%%%%%%%%%%%%%%%%%%%%%%%%%%%%%
%%%%%%%%%%%%%%%% Summary %%%%%%%%%%%%%%%%%%%%%%%
%%%%%%%%%%%%%%%%%%%%%%%%%%%%%%%%%%%%%%%%%%%%%%%%%%
\section{Summary}
\label{Summary}

A new silicon-strip detector was developed for the sPHENIX experiment at RHIC. The silicon ladder consisted of silicon-strip sensors, FPHX chips, HDIs, and a high thermally conductive carbon-fiber stave. The ladder was designed to be as thin as $X/X_0=1.14$\%. The bus extender and the conversion cables were developed to transmit signals from the ladder to the ROC board. The 1.11 m long bus extender cable employs the novel low attenuation LCP as the dielectric material of the FPC. The conversion cable employs $\mu$-coaxial harnesses to secure flexibility in three dimensions in order to connect the bus extender end to the input/output ports of the ROC board without introducing any stress at the connection. 

The result of the study indicated that both the bus extender and the conversion cables are sufficiently radiation hard against estimated radiation dose for three years of the INTT operation in the sPHENIX experiment. In contrast, the exposed dose for TLK2711 chips of the ROC board throughout five years of the FVTX operation in the PHENIX experiment caused some bit errors to start to appear. Any chips showing these symptoms were replaced with new chips.
The ROCs are installed relatively far away from the collision point for INTT operation in sPHENIX as opposed to FVTX in PHENIX.  Thus, the radiation dose for the ROC boards in three years of INTT operation in sPHENIX is estimated to be relatively moderate, i.e. 1/6 of that of the FVTX in PHENIX.

%%%%%%%%%%%%%%%%%%%%%%%%%%%%%%%%%%%%%%%%%%%%%%%%%%%%%%%%%%%%%%%%%%%%%%
\section*{Acknowledgments}
%%%%%%%%%%%%%%%%%%%%%%%%%%%%%%%%%%%%%%%%%%%%%%%%%%%%%%%%%%%%%%%%%%%%%%

We are grateful to the sPHENIX collaboration for their support and help during the various phases of the INTT ladder and the readout cable developments. We thank the staff of physics and instrumentation divisions in Brookhaven National Laboratory. We thank FVTX experts for providing us various documentations and advices.
We thank the staff of National Institutes for Quantum Science and Technology and RANS facility for supporting the radiation exposure to cable samples. We thank the staff of Tokyo Metropolitan Industrial Technology Research Institute for various measurements to evaluate the signal transmission performance of readout cables.
We thank Mr. H. Yanami, Mr. Y. Yanami, Mr. H. Takahashi, and Mr. K. Matsumoto from Print Electronics Laboratory Co., Ltd., and Mr. T. Miyazaki from Taiyo Manufacturing Co., Ltd., for their tremendous support for the R\&D and production of the BEX. We thank Mr. Kimura from Kimuraya Co., Ltd., and Mr. T. Yoshikawa from Asuka Co., Ltd. for R\&D and production of the stave.
Lastly, we are also grateful to Mr. D. Yanagawa and Mr. H. Hoshiya from HAYASHI-REPIC Co., Ltd. for their tremendous supports on engineering aspects of the INTT ladder and readout cables from the beginning of R\&D.   

We acknowledge support from the Ministry of Education, Culture, Sports, Science, and Technology and the Japan Society for the Promotion of Science (Japan); the collaborative development fund of Tokyo Metropolitan Industrial Technology Research Institute, Office of Nuclear Physics in the Office of Science of the Department of Energy, the National Science Foundation (U.S.A.); National Science and Technology Council and the Ministry of Education (Taiwan); and the National Research Foundation of Korea (NRF), Ministry of Science and ICT (MSIT) (No. 2018R1A5A1025563) (Republic of Korea).

%%%%%%%%%%%%%%%%%%%%%%%%%%%%%%%%%%%%%%%%%%%%%%%%%%
%%%%%%%%%%%%%%%% Reference %%%%%%%%%%%%%%%%%%%%%%%
%%%%%%%%%%%%%%%%%%%%%%%%%%%%%%%%%%%%%%%%%%%%%%%%%%
%\input{INTT_Reference.tex}
\bibliographystyle{cas-model2-names}
\bibliography{INTT_Ladder_NIM_main}

\end{document}